\documentclass[submission,journal]{IEEEtran}
\usepackage{epsfig}
\usepackage{amssymb}
\usepackage{float}
\usepackage{array}
\usepackage[nodots]{numcompress}
\usepackage{graphicx} 
\usepackage{subfigure}
\usepackage{epsfig} 
\usepackage{amsmath} 
\usepackage{amssymb}  
\usepackage{cite}
\usepackage{url}
\usepackage[OT1]{fontenc} 
\usepackage{float}
\usepackage{graphicx}
\usepackage[misc]{ifsym}
\usepackage[justification=centering]{caption}

\usepackage[ruled,linesnumbered]{algorithm2e}

\usepackage{booktabs}
\usepackage{multirow}
\usepackage{siunitx}
\usepackage{amsmath}
\usepackage{tabularx}
\usepackage{algpseudocode}
\algnewcommand\algorithmicswitch{\textbf{switch}}
\algnewcommand\algorithmiccase{\textbf{case}}
\algnewcommand\algorithmicassert{\texttt{assert}}
\algnewcommand\Assert[1]{\State \algorithmicassert(#1)}%
\algdef{SE}[SWITCH]{Switch}{EndSwitch}[1]{\algorithmicswitch\ #1\ \algorithmicdo}{\algorithmicend\ \algorithmicswitch}%
\algdef{SE}[CASE]{Case}{EndCase}[1]{\algorithmiccase\ #1}{\algorithmicend\ \algorithmiccase}%
\algtext*{EndSwitch}%
\algtext*{EndCase}%
\newcommand{\tabincell}[2]{\begin{tabular}{@{}#1@{}}#2\end{tabular}}
\newcolumntype{N}{@{}m{0pt}@{}}

\newtheorem{myDef}{Definition} 

\usepackage{diagbox}
\newcolumntype{N}{@{}m{0pt}@{}}

\begin{document}
%
\title{What happens to a ToF LiDAR in fog?}
%
%
%

\author{You Li$^{1}$,
        Pierre Duthon$^{2}$,
        Mich\`ele Colomb$^{2}$,
        Javier Ibanez-Guzman$^{1}$
\thanks{$^{1}$You Li and Javier Ibanez-Guzman({\tt\small you.li@renault.com, javier.ibanez-guzman@renault.com}) are with the research department of RENAULT S.A.S, 1 Avenue du Golf, 78280 Guyancourt, France. }
      \thanks{$^{2}$Pierre Duthon ({\tt\small pierre.duthon@cerema.fr}) and Mich\`ele Colomb ({\tt\small michele.colomb@cerema.fr}) are with CEREMA, Equipe-projet STI, 8-10 rue Bernard Palissy, CEDEX 2, F-63017 Clermont-Ferrand, France.}}%

\maketitle

\begin{abstract}
  By transmitting lasers and processing laser returns, LiDAR (light detection and ranging) perceives the surrounding environment through distance measurements. Because of high ranging accuracy, LiDAR is one of the most critical sensors in autonomous driving systems. Revolving around the 3D point clouds generated from LiDARs, plentiful algorithms have been developed for object detection/tracking, environmental mapping, or localization. However, a LiDAR's ranging performance suffers under adverse weather (e.g. fog, rain, snow etc.), which impedes full autonomous driving in all weather conditions. This article focuses on analyzing the performance of a typical time-of-flight (ToF) LiDAR under fog environment. By controlling the fog density within CEREMA Adverse Weather Facility\footnote{\url{https://www.cerema.fr/fr/innovation-recherche/recherche/projets/adverse-weather-environmental-sensing-system-dense}}, the relations between the ranging performance and fogs are both qualitatively and quantitatively investigated. Furthermore, based on the collected data, a machine learning based model is trained to predict the minimum fog visibility that allows successful ranging for this type of LiDAR. The revealed experimental results and methods are helpful for ToF LiDAR specifications from automotive industry.     
\end{abstract}

\IEEEpeerreviewmaketitle

\section{Introduction}
As an active sensor, LiDAR (light detection and ranging) illuminates the surroundings by emitting lasers. The reflected laser pulses are then detected by certain photodetectors, such as APD (avalanche photodiode) or SPAD (single-photon avalanche diode). Range measurements are acquired by processing the laser returns with regard to the emitted lasers. Knowing the pose the LiDAR allows to calculate the 3D Cartesian coordinates from 1D ranges. Comparing with camera and radar, LiDAR is much better in ranging accuracy and precision \cite{PerceptionJFR2008}. Therefore, LiDAR is always regarded as a critical sensor to assure safety for high level autonomous vehicles \cite{youliSPM2020}. In DARPA Grand Challenge 2007 -- a milestone in autonomous driving history, all the top 3 teams were equipped with multiple LiDARs. Applications of LiDAR in autonomous driving can be divided into two categories: 1) perception, such as object detection, tracking and recognition\cite{Stefan2018, edouardIV19}; 2) localization and mapping\cite{loam2014, edouardITSC2018}.

However, most of the applications assume that LiDARs are always working within perfect environments, ignoring the impact of adverse conditions such as fog, rain or snow. In literature, the researches on this subject (e.g. \cite{jfr2019, imran2018, daimler2018}) are insufficient. With the fast progress of autonomous driving systems, the impacts of adverse weather on LiDARs become non-negligible for deploying full self-driving cars.

\begin{figure}
\centering
\includegraphics[width = 0.4\textwidth]{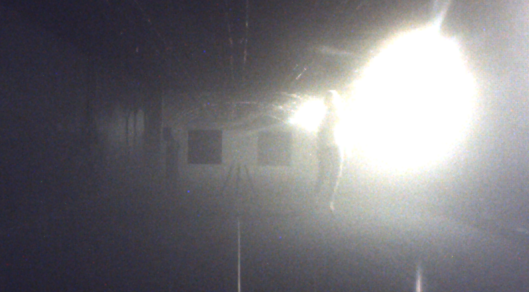}
\caption{A scenario of testing LiDAR performance under fog in CEREMA Adverse Weather Platform.}
\label{fig::fog_example}
\end{figure}

In this paper, we present a performance analysis and modeling of a popular ToF LiDAR, Velodyne UltraPuck\footnote{\url{https://velodynelidar.com/vlp-32c.html}}, under well-controlled artificial fog environments. Fig. \ref{fig::fog_example} shows a testing scenario. Time-of-flight (ToF) LiDAR, one of the most popular LiDAR types, computes the time differences between the transmitted and received lasers. Another category of LiDAR is FMCW (frequency modulated continuous wave), which measures the range and velocity based on the doppler effect. ToF LiDAR's structure is simple and is quite mature in manufacture. FMCW LiDAR is much more expensive and not yet popular in the automotive domain. Therefore, in this paper, we choose ToF LiDAR for performance test. The contributions of this paper are twofold: (1) At first, comparing with some works in literature, very detailed experimental results are both qualitatively and quantitatively analyzed. (2) Based on the collected data, a machine learning based model is trained to predict when a LiDAR would fail under fog condition. As far as the authors' knowledge, this is the first work to quantitatively analyze and model a ToF LiDAR's performance under fog conditions in a data-driven approach.           

This paper is organized as follows: Sec \ref{sec::review} reviews related researches. Theoretical model of a ToF LiDAR and the factors impacting a ToF LiDAR's ranging capability under adverse weather are discussed in Sec \ref{sec::theory}. Then, the experiments are described in Sec. \ref{sec::experiment}, and the results are analyzed in Sec. \ref{sec::analyze}. At last, based on the collected data, a machine learning based method is proposed in Sec. \ref{sec::gpr} to model the performance of the tested LiDAR in fog environment. 

\section{Literature Review} \label{sec::review}
\begin{figure*}[t]
\centering
\includegraphics[width = 0.8\textwidth]{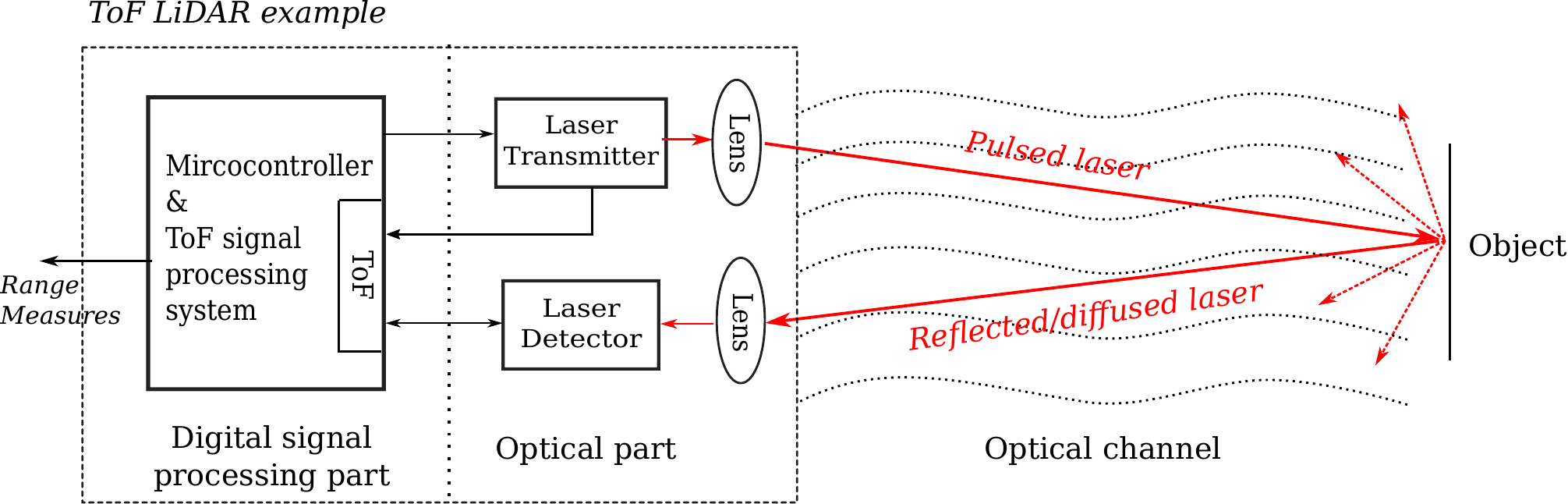}
\caption{An example of a ToF LiDAR system.}
\label{fig::lidar_system}
\end{figure*}

In literature, there are a few studies on the impacts of adverse weather on LiDARs. Some researchers tried to develop a theoretical model to simulate the behavior of LiDAR under adverse weather. For instance, in \cite{polland2014}, the authors compared the range degradation of a 905nm ToF LiDAR with a 1050nm one due to adverse conditions. Atmospheric extinction coefficients and reflectances of various materials are deducted from theoretical models to infer the results of LiDARs. A simplified model of LiDAR's performance in rain conditions is proposed in \cite{electronics2019} for a simulation software. However, both of these two works were not verified by real experiments. \cite{BMW2011} modeled the impact of bad weather on a 905nm ToF LiDAR based on Mie scattering theory. Although the developed model was verified through real experiments, only several rough visibilities of adverse weather are tested due to facility restrictions.

Although theoretical models represent the physical characters of LiDAR, they are always built on assumptions and simplifications rarely being held on real environments. Therefore, some researches emphasized on empirical evaluation. In \cite{jfr2019}, a radar and two LiDARs (SICK and Riegl) are tested in rain, mist and dust conditions. Radar is found to be more robust than the tested LiDARs in such environments. Range errors of LiDARs are estimated as well. \cite{assessment2008} assessed four different LiDARs under visibility reduced environments with water vapor or smoke. In \cite{imran2018}, various fog conditions are created to test Veldoyne HDL-64E, a LiDAR of 905nm wavelength. A metric named SSIM (Structural Similarity Index Measurement) is used to measure the impact of fog attenuation, w.r.t the visibility of fog. \cite{measurement2017} tried to quantify the influence of rain to Velodyne VLP16. Range and intensity changes are both investigated through field tests. However, the rain conditions are not well measured: the utilized weather data is too general to quantitatively analyze the LiDAR performance w.r.t rain density. \cite{jlt2013} investigated the fog and smoke attenuation for NIR (near infrared) wavelength lasers under a 5.5m long atmospheric chamber, for the purpose of optical communication. The distances tested ($<6m$) is insufficient for LiDAR applications and atmospheric attenuation is just one of the factors impact LiDAR performance under adverse weather. Within an EU project DENSE (aDverse wEather eNvironment Sensing systEm)\footnote{\url{https://www.dense247.eu/home/}} aiming to develop perception sensors which can work under bad weather conditions, \cite{matti2018}, \cite{matti2019} tested and benchmarked various range sensors within a well-controlled fog and rain facility at CEREMA. Also within the same facility, \cite{daimler2018} quantitatively benchmarked a Velodyne HDL64 LiDAR and a IBEO Lux4 LiDAR, which are both in 905nm wavelength. 

\section{Theoretical Model of ToF LiDAR and Adverse Weather Impacts}\label{sec::theory}
In this section, we summarize the principle of a ToF LiDAR and the factors impacting its performance under adverse weather. 
\subsection{Principle of a ToF LiDAR} \label{sec::lidar_principle}
As the most popular LiDAR category, ToF (time-of-flight) LiDARs measure distances by calculating the time difference between emitted laser pulses and the diffused or reflected lasers from obstacles. The equation of a ToF LiDAR is given as:
\begin{equation}
  R = \frac{1}{2n}c\Delta t
  \label{eq::tof}
\end{equation}
where $R$ is the measured range, $c$ is the light speed, $n$ is the index of refraction of the propagation medium (approximately 1 for air). $\Delta t$ is the time gap between the transmitted laser and received laser.

A typical ToF LiDAR system comprises three parts: \textit{transmitter, receiver, time control and signal processing circuits}, as shown in Fig. \ref{fig::lidar_system}. Driven by the microcontroller, a pulsed laser is transmitted through certain transmission medium, air for instance, to illuminate the surroundings. When the emitted laser hits on an object, diffused or reflected laser returns are captured by the receiver's optical system and are transformed into electrical signals by photodetectors, such as APD (Avalanche Photon Diode). This process can be summarized by LiDAR's power model: 

\subsubsection{Power model}
The power of a received laser return at distance $R$ can be modeled as (\cite{IntroLiDAR2005,BMW2011}):
\begin{equation}
  P(R) = E_p\frac{c\eta A}{2R^2} \cdot \beta \cdot T(R)
  \label{eq::lidar}
\end{equation}
Where $E_p$ is the total energy of a transmitted pulse laser, $c$ is light speed. $A$ represents receiver's optical aperture area. $\eta$ is the overall system efficiency. $\beta$ is the reflectivity of the target's surface, which is decided by surface properties and incident angle. In a simple case of Lambertian reflection with a reflectivity of $0<\Gamma<1$, it is given by:
\begin{equation}
  \beta = \Gamma/\pi
\end{equation}



The final part $T(R)$ denotes the transmission loss through the transmission medium, which is given by:
\begin{equation}
T(R) = exp(-2\int_0^{R}\alpha(r)dr)  
\end{equation}
$\alpha(r)$ is the extinction coefficient of the transmission medium. The extinction is due to the particles within the transmission medium that would scatter and absorb the laser.

\begin{figure*}[t]
  \centering
\subfigure[Overall structure of CEREMA's platform]{
\includegraphics[width = 0.52\textwidth]{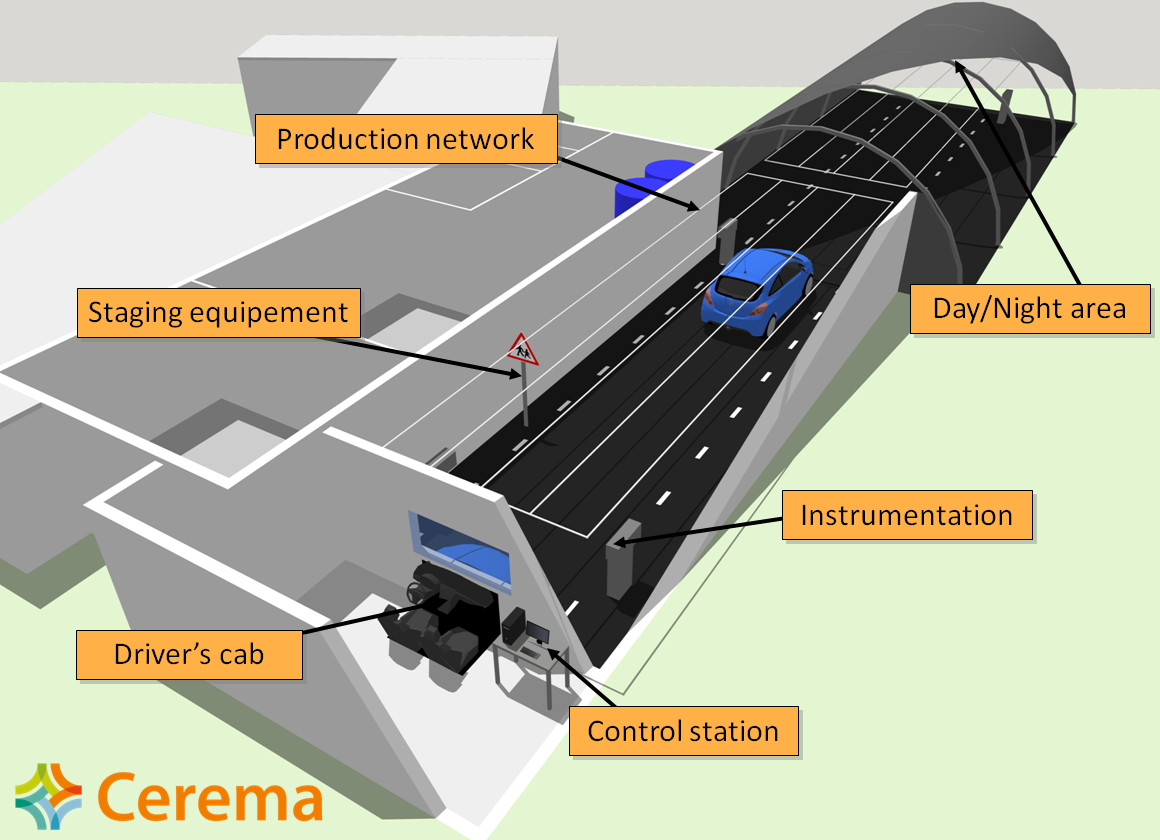}
}
\subfigure[Instruments in CEREMA's platform.]{
\includegraphics[width = 0.37\textwidth]{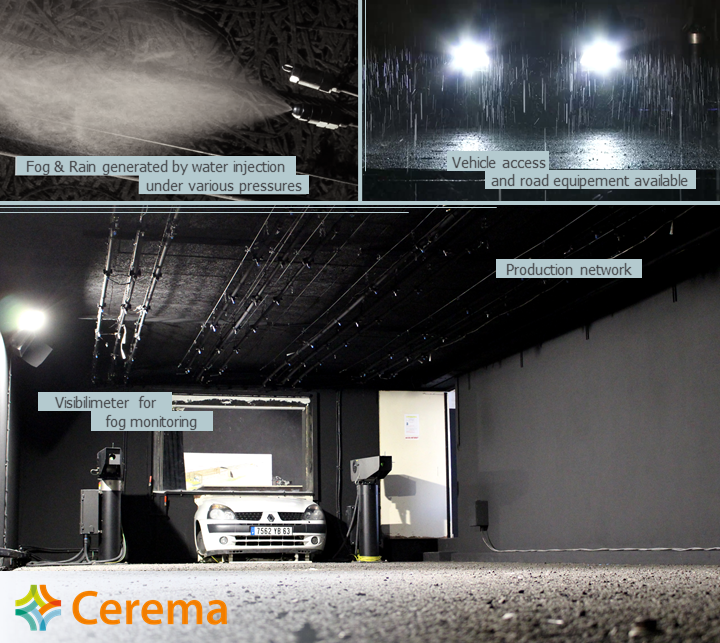}
}
\caption{CEREMA's Adverse Weather facility}
\label{fig::cerema}
\end{figure*}

\subsubsection{Pulse detection}
After transforming the laser returns into electrical signals, a signal processing unit detects the received laser signal from background noises. Finally, the time difference $\Delta t$ is got from the detected return signal and the range $R$ is calculated as in Eq. \ref{eq::tof}. Eq. \ref{eq::lidar} reveals that, the energy of received laser decrease quadratically with regard to the distance. Simply increasing the power of the transmitted laser is infeasible due to eye-safety restrictions, such as IEC 60825\cite{iec60825}. Therefore, advanced signal processing algorithms capable of detecting the true return signal in low SNR (signal-to-noise ratio) are required. To increase the SNR, a low-pass filter or a band-pass filter is usually applied inside the signal processing circuits \cite{AP1971}. A thresholding algorithm is applied on the raw data to detect the true return signal. Adaptive thresholding methods capable of learning the statistics of the background noises are widely applied, such as the well-known constant false alarm rate (CFAR) detector \cite{CFAR2016}, or the methods in \cite{patent2012} or \cite{adaptSPAD2018}.

\subsection{Influences of Adverse Weather} \label{sec::influence}
From the short review of a ToF LiDAR's principle, it can be inferred that the adverse weather, such as fog or rain, enlarges the transmission loss $T(R)$ and hence leads to lower received laser power $P(R)$, which fails the following signal processing step. In fact, LiDAR's performance degrades due to the change of extinction coefficient $\alpha$\cite{Bjorn1998} and target's reflectivity $\beta$\cite{wet1988}:

\begin{itemize}
\item Impact on extinction coefficient $\alpha$: The droplets in the fog or the rain would absorb or scatter the near infrared laser \cite{polland2014}. The severity depends on the water content percentage, droplet size distribution \cite{BMW2011}, etc.    
\item Impact on surface reflectivity $\beta$. A wet surface always looks "darker" than dry surface \cite{wet1988}, because a thin film of liquid on an obstacle's wet surface lead to weaker diffuse reflection. The decreased surface reflectivity $\beta$ hence leads to a reduced maximum detection range in adverse weather. 
\end{itemize}




\section{Experiments}\label{sec::experiment}
Apart from theoretical analysis, we are interested to know the empirical results of how a ToF LiDAR perform under varying fog conditions. We realized the tests within CEREMA Adverse Weather facitliiy\footnote{\url{https://www.cerema.fr/fr/innovation-recherche/innovation/offres-technologie/plateforme-simulation-conditions-climatiques-degradees}} -- a center in Europe generating controlled adverse weather conditions, as shown in Fig. \ref{fig::cerema}. In our experiments, the popular Velodyne UltraPuck was chosen because of its wide applications in autonomous vehicles. A technical summary of this sensor is shown in Tab. \ref{tab::sensors}. Under fog environment, various targets were put at different distances with regard to the LiDAR, and the correspondent LiDAR measures are recorded for further analysis. 
\begin{table}[h]
  \centering
  \begin{tabular}{r|l}
\toprule
          &\tabincell{l}{Velodyne UltraPuck} \\\hline
    Max range &    200m ($80\%$ reflectivity)              \\\hline
    Range accuracy &   5cm              \\\hline
    Horizontal FOV &  $360^\circ$      \\\hline
    Vertical FOV &  $40^\circ$       \\\hline
    Horizontal angular resolution & $0.05^\circ \sim 0.2^\circ$ \\\hline
    Vertical angular resolution & $0.33^\circ$ (min)      \\\hline
    Laser wavelength & 903nm                 \\\hline
    Max scan rate    & 20hz                 \\
\bottomrule    
\end{tabular}
\caption{A summary of Velodyne UltraPuck}
\label{tab::sensors}
\end{table}

\subsection{CEREMA's Adverse Weather Facility}
The CEREMA Adverse Weather Platform, was developed to investigate all transport systems that could be affected by adverse conditions, including fog and rain \cite{Colomb2008, DuthonTRA2016}. It allows to reproduce various scenarios, as detection of vulnerable road users or fixed obstacles, in clear conditions, night conditions, and with various ranges of fog and rain precipitations on a total length of 30 meters (Fig. \ref{fig::cerema} (a)). Dedicated to research and development, it is also open to private companies looking for a testing facility with controlled conditions. It has been used for years in partnership or collaborative projects in order to investigate various scientific topics, as  humans’ perception in adverse conditions, vision systems capabilities in fog or rain conditions \cite{Bernardin2014, Pinchon2016} or computer vision algorithms for objects detection \cite{Dahmane2016, Bijelic2019}. The physical characteristics of rain and fog produced in the platform are described in a recent study on LiDARs performances in fog and rain \cite{matti2018}.

This platform has got a high-level instrumentation to evaluate the performance of perceptual sensors for autonomous vehicles in adverse conditions. Some weather instruments are dedicated to characterizing the atmosphere in fog or rain conditions (as shown in Fig. \ref{fig::cerema} (b)):

\begin{itemize}
\item Transmissometer, for meteorological visibility in fog from 5 to 1000 $m$ with 1HZ recording,
\item Optical granulometer, for fog droplet size distribution from 0.4 to 40 $\mu m$ with 1 minute step recording, 
\item Rain gauge and a spectro-pluviometer for rainfall rate from 0.001 to 1200 $mm/h$ with 1HZ recording.
\end{itemize}

\subsection{Test Methodology}
\subsubsection{Artificial fog and visibility data}
\begin{figure}[t]
  \includegraphics[width = 0.45\textwidth]{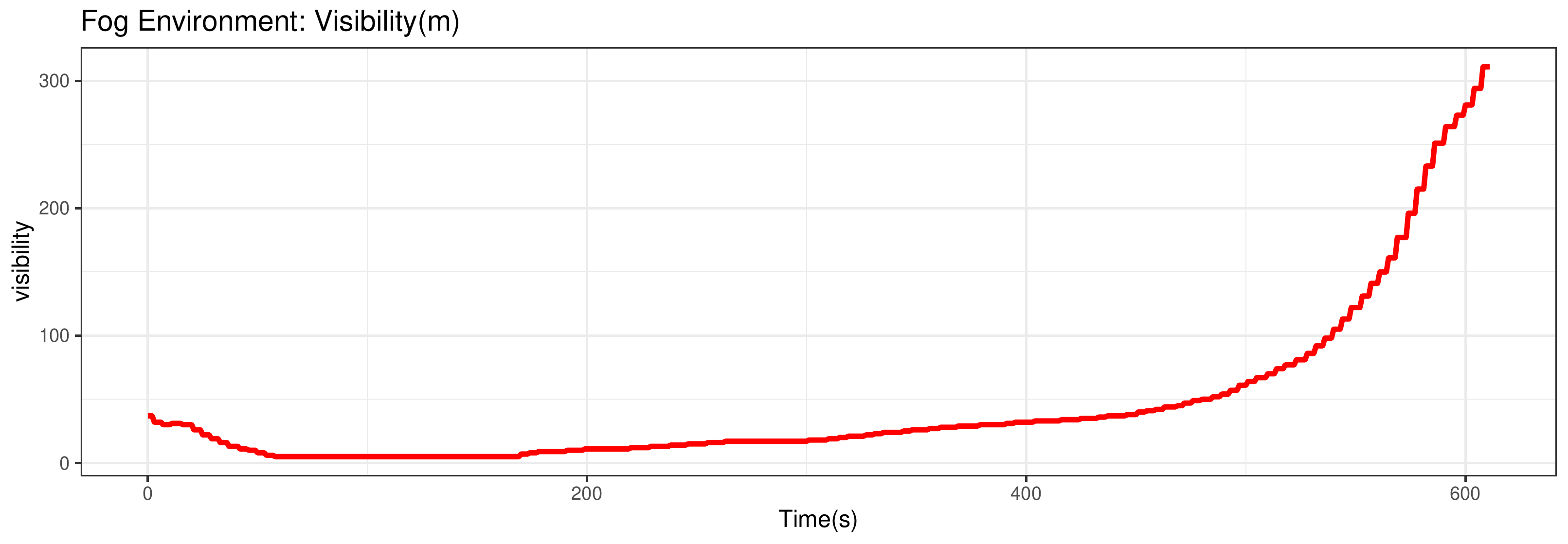}
\caption{A sample of visibility recording in an artificial fog: the visibility reaches almost 10m and then gradually increases to more than 300m due to dissipation.}
\label{fig::meteo}
\end{figure}

As shown in Fig. \ref{fig::cerema} (b), nozzles are distributed inside the chamber. These nozzles used at high pressure are capable of mechanically producing water droplets that are similar in size to a natural fog. As the atmosphere is not saturated enough with water, the droplets will gradually evaporate, we call it dissipation. Thus, by precisely monitoring in real time the quantity of water injected into the test chamber, it is possible to regulate the meteorological visibility, thanks to the usage of transmissometers (Fig. \ref{fig::cerema} (b)).
In the beginning of each test, we generate a dense fog reaching the minimum available meteorological visibility of 10m. Then, we let the fog gradually dissipate. The fog dissipation leads to an increase of meteorological visibility, until the air becomes clear. (From a meteorological point of view, there is no fog if the meteorological visibility is more than 1000m. But a french road standard considers that there is no fog when the meteorological visibility is more than 400m). Fig. \ref{fig::meteo} shows a real example of the visibility recordings during a test of around 600 seconds. The change of visibility reflects the change of fog’s density.

\begin{figure*}[t]
  \centering
  \subfigure[Testing setup inside the fog chamber of CEREMA]{
\includegraphics[width = 0.5\textwidth]{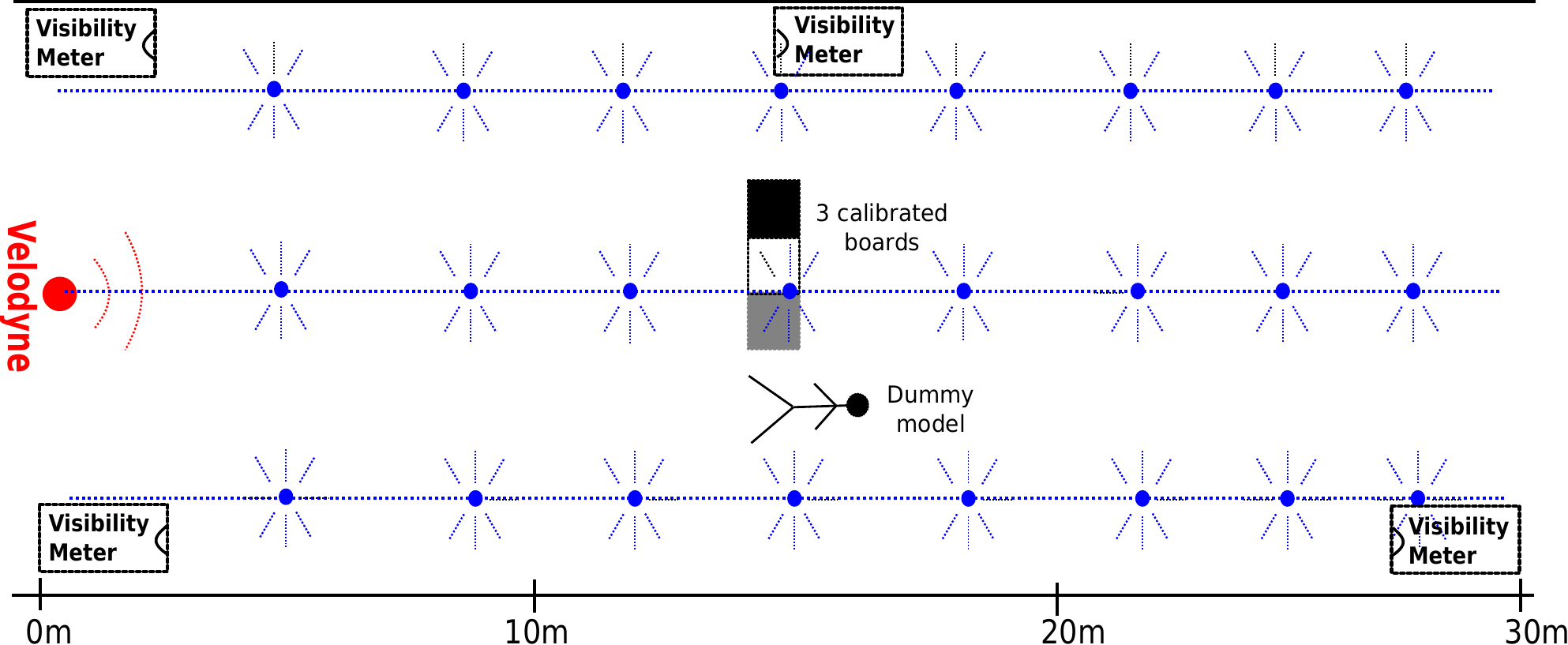}
}
\subfigure[Top: used targets (3 calibrated boards, vehicle, a dummy model and 2 traffic signs). Bottom: Velodyne UltraPuck on the table and an example scenario]{
\includegraphics[width = 0.3\textwidth]{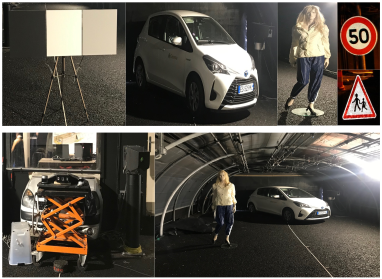}
}
\subfigure[LiDAR measures for all the targets (at 15m)]{
\includegraphics[width = 0.8\textwidth]{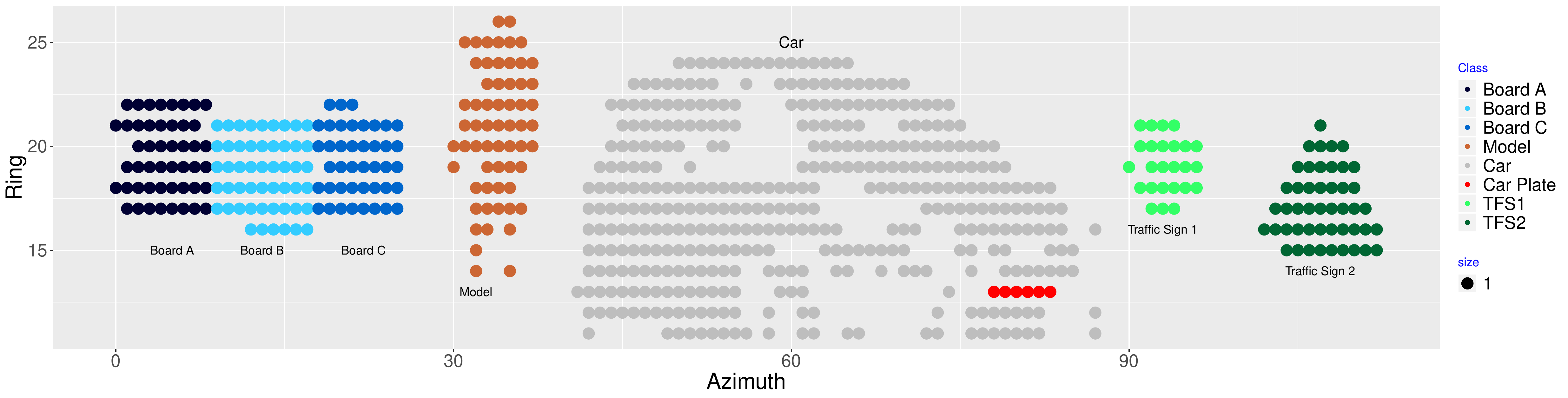}
}
\subfigure[Reflectivity distributions of diffuse targets (at 15m in clear weather, fitted by a Gaussian distribution) ]{
\includegraphics[width = 0.4\textwidth]{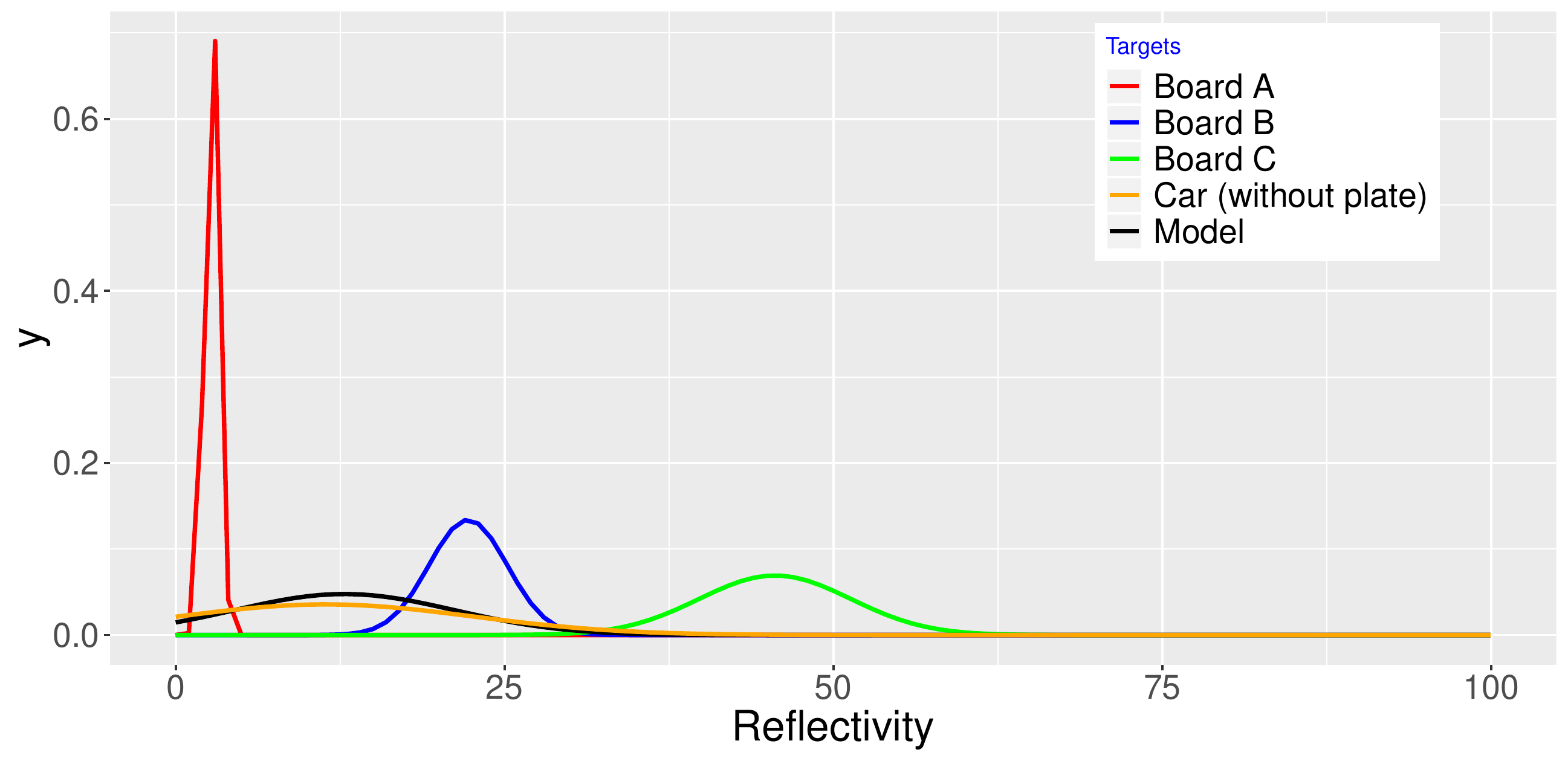}
}
\subfigure[Reflectivity distributions of retro-reflected targets (at 15m in clear weather, fitted by a Gaussian distribution)]{
\includegraphics[width = 0.4\textwidth]{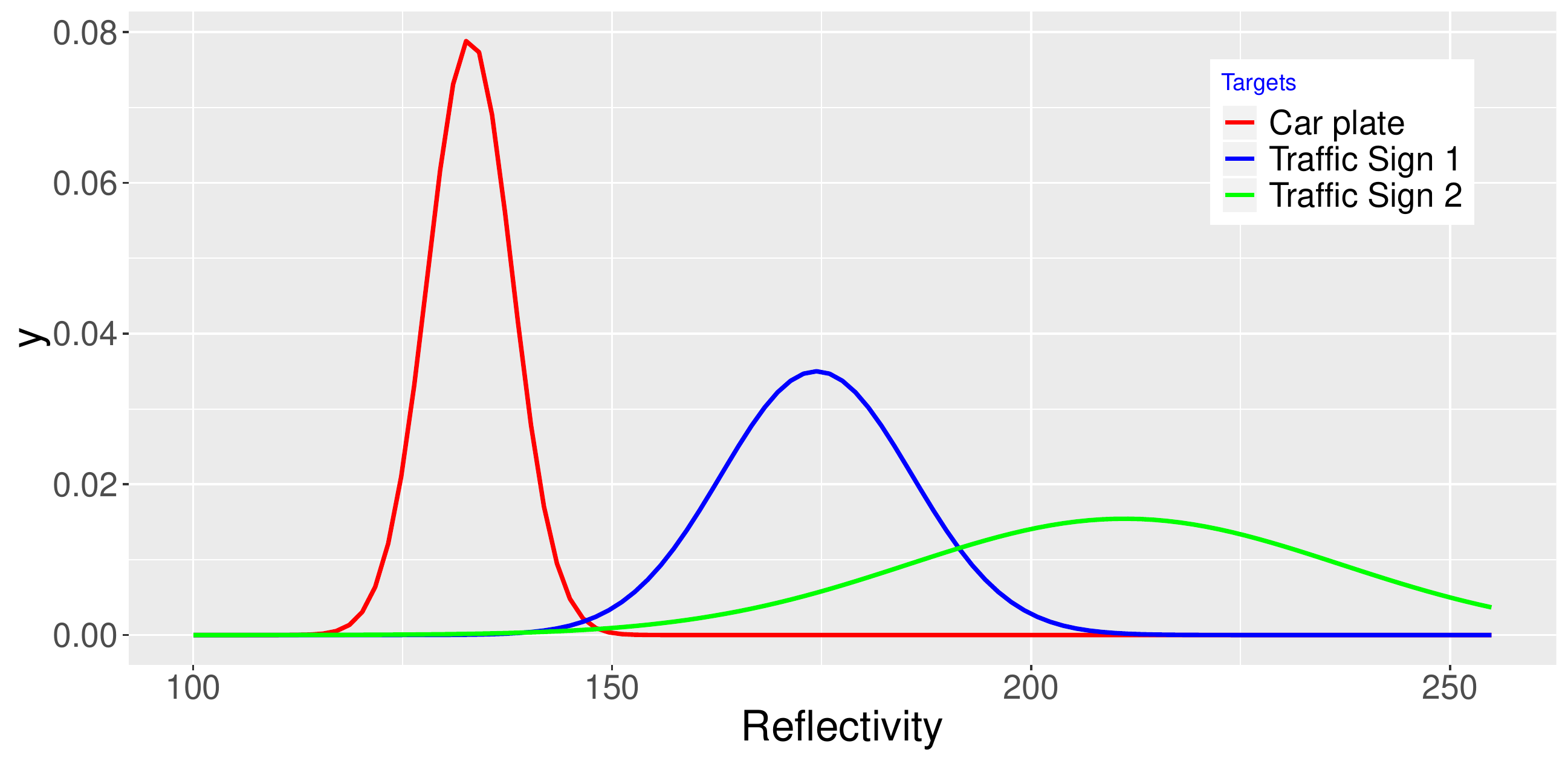}
}
\caption{Scenarios and targets in testing}
\label{fig::method}
\end{figure*}

\subsubsection{Targets}
Fig. \ref{fig::method} (a) sketches a setup of our tests within the fog chamber. The Velodyne UltraPuck is put on a height-adjustable table (as shown in Fig. \ref{fig::method} (b)). It works at 10HZ in the strongest return model. Several typical road targets are put in the platform with various distances. Those targets are: (1) three well-calibrated Zenith Polymer boards (A, B, C) with reflectivities A: $5\%$, B: $50\%$ and C: $90\%$, (2) a dummy model, (3) a car and (4) two traffic signs (TFS1 and TFS2). The used targets and correspondent LiDAR measures are shown in Fig. \ref{fig::method} (b) and (c). Velodyne UltraPuck returns a calibrated reflectivity byte (0-255) for each range measure, enabling distinguishment of retro-reflectors (e.g. road sign, license plate) from diffuse reflectors (e.g. road, tree trunk). The measured reflectivity has either:
\begin{itemize}
\item a value between 0 to 100 for diffuse reflectors, an approximation of reflectivity based on the ratio of emitted and received laser power. 
\item a value between 101 to 255 for retro-reflectors, characterizes a continuum from a dirty or imperfect retro-reflector to a more robust retro-reflector at an ideal angle.
\end{itemize}

Within the utilized targets, the three calibrated boards, the dummy model, and the car except plate region belong to diffuse reflectors. Fig. \ref{fig::method} (d) shows fitted Gaussian distributions of reflectivities for these diffuse targets at 15m, (e) demonstrates the fitted reflectivity distributions of three retro-reflectors (car plate and two traffic signs). Three calibrated boards obviously distinguish each other. The dummy model and car without plate are similar: both of the two targets' reflectivities range between 0 to 35.             

\begin{figure*}[t]
  \centering
\subfigure[Ground truth of a test scenario (Model, 3 boards, R=15m)]{
\includegraphics[width = 0.46\textwidth, height=3cm]{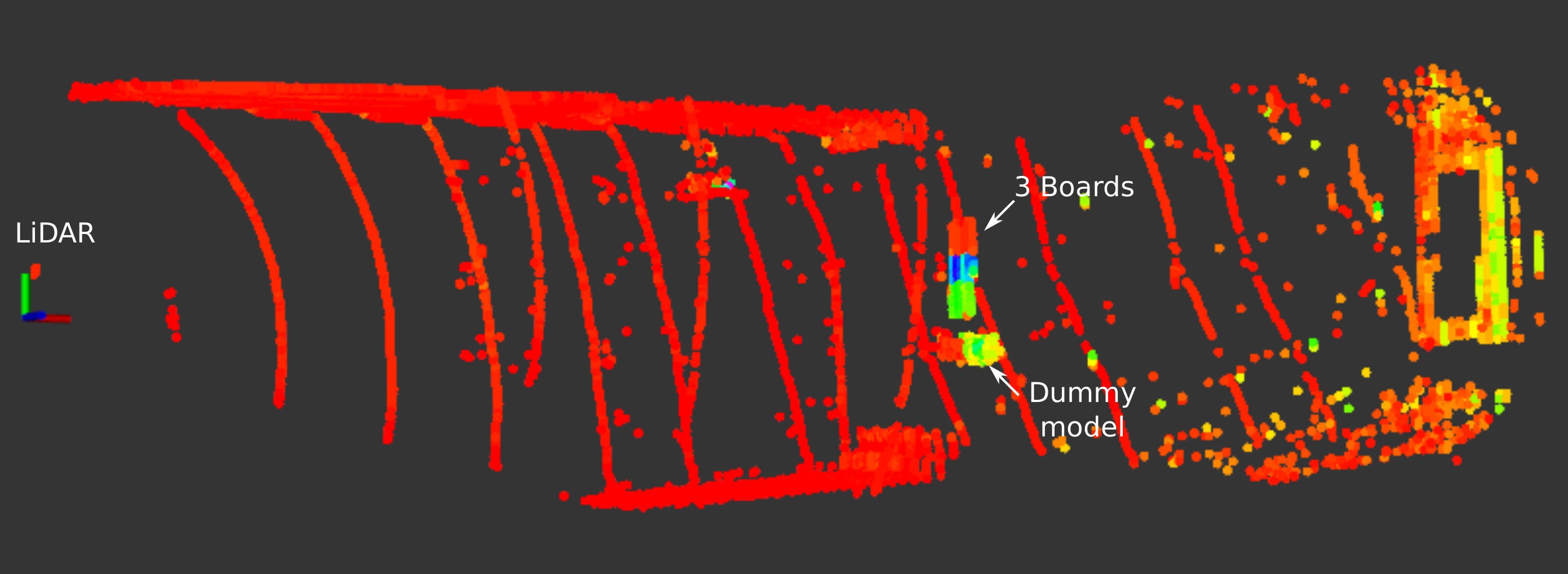}
}
\subfigure[All the LiDAR measures (Model, 3 boards, R=15m, V=55m)]{
\includegraphics[width = 0.46\textwidth, height=3cm]{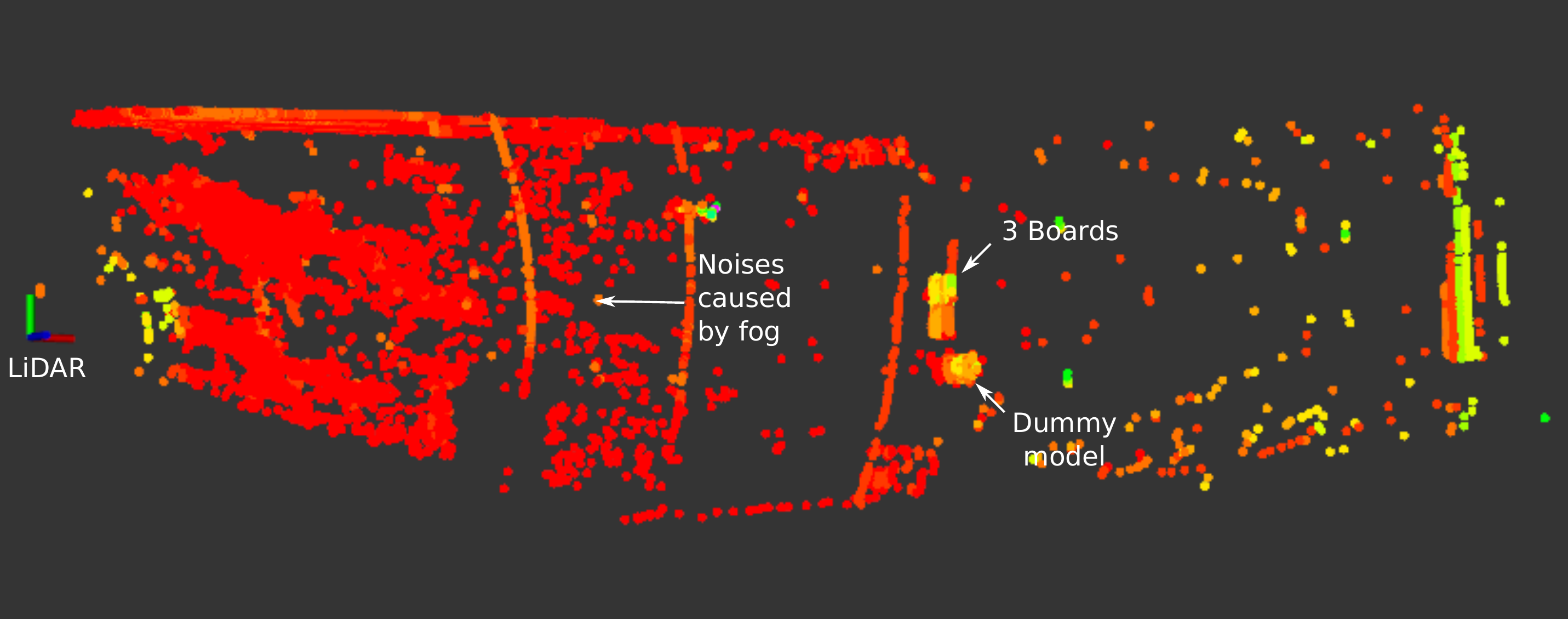}
}

\subfigure[LiDAR measures of targets (Model, 3 boards, R=15m V=40m)]{
\includegraphics[width = 0.22\textwidth,height=3.5cm]{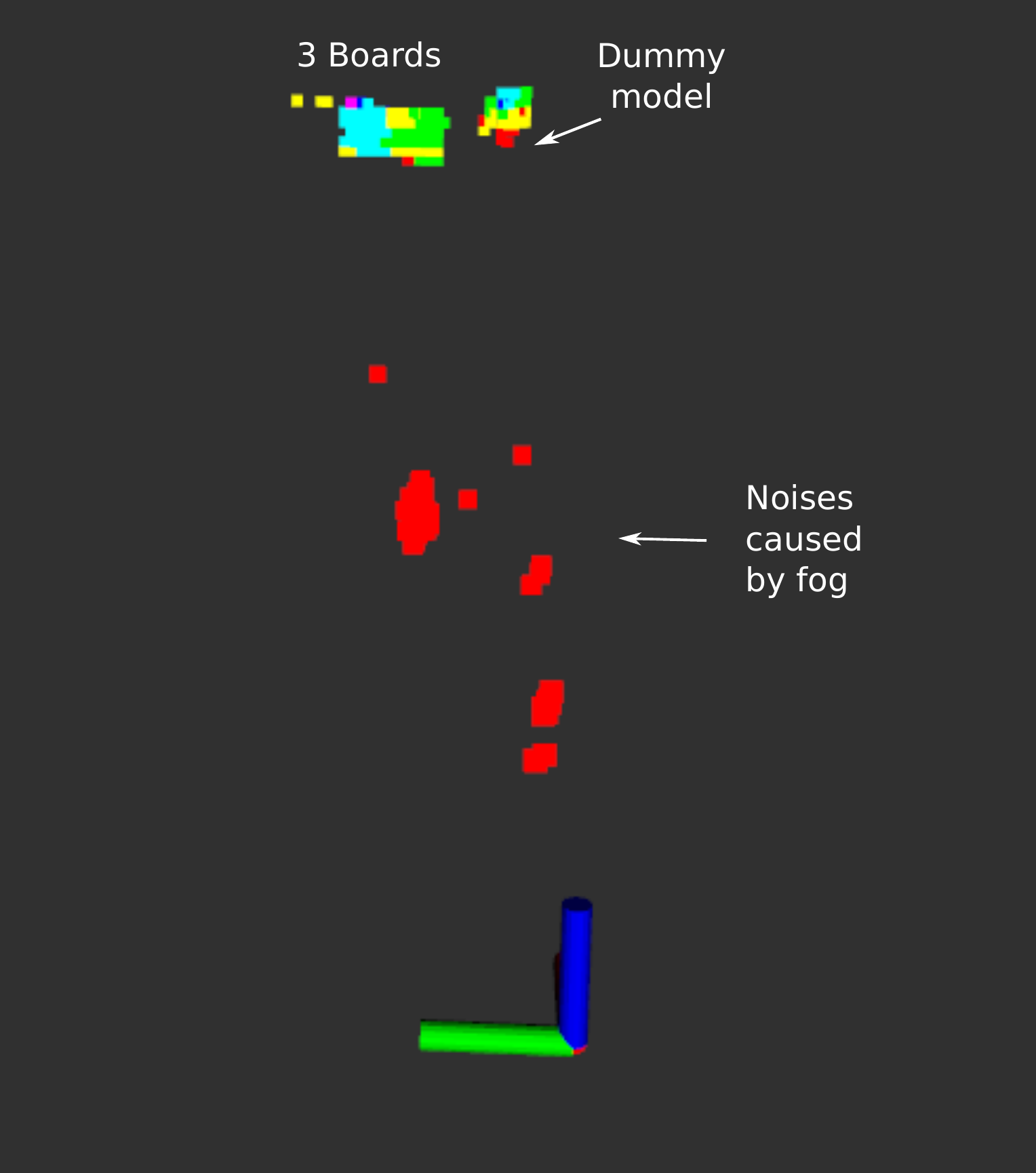}
}
\subfigure[LiDAR measures of targets (Model, 3 boards, R=15m V=80m)]{
\includegraphics[width = 0.22\textwidth,height=3.5cm]{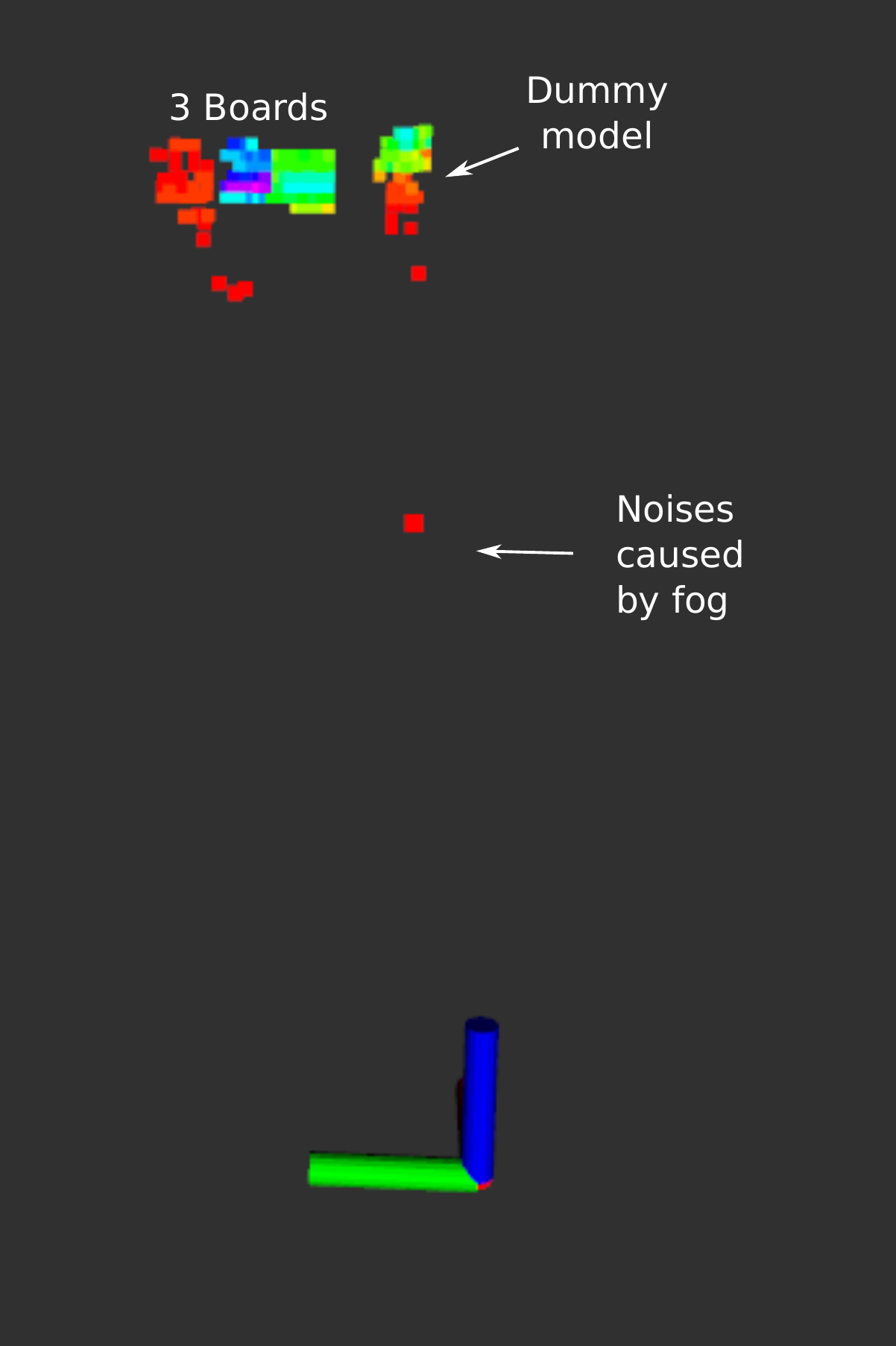}
}
\subfigure[LiDAR measures of targets (Model, 3 boards, R=20m, V=40m)]{
\includegraphics[width = 0.22\textwidth, height=3.5cm]{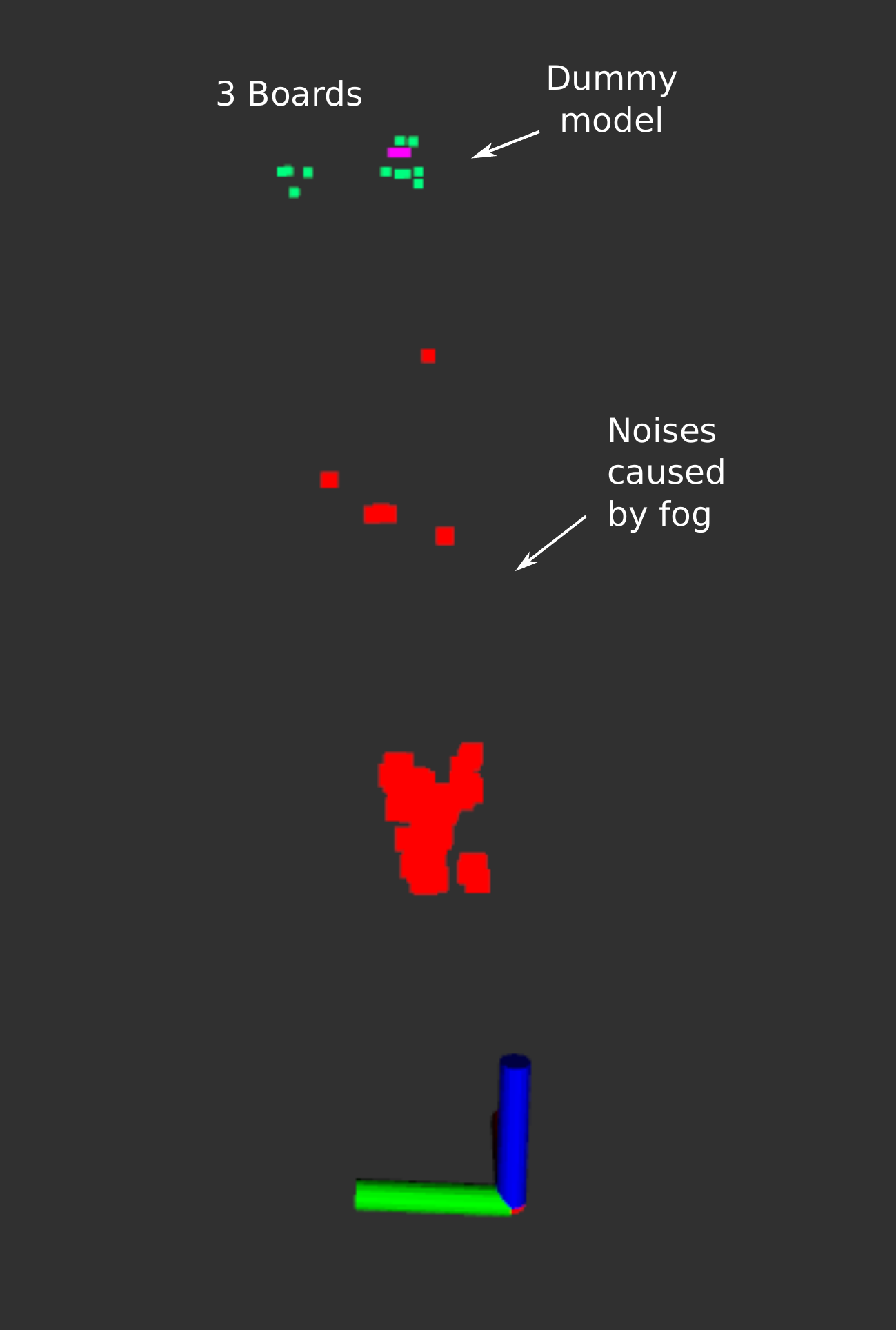}
}
\subfigure[LiDAR measures of targets (Model, 3 boards, R=20m, V=80m)]{
\includegraphics[width = 0.22\textwidth, height=3.5cm]{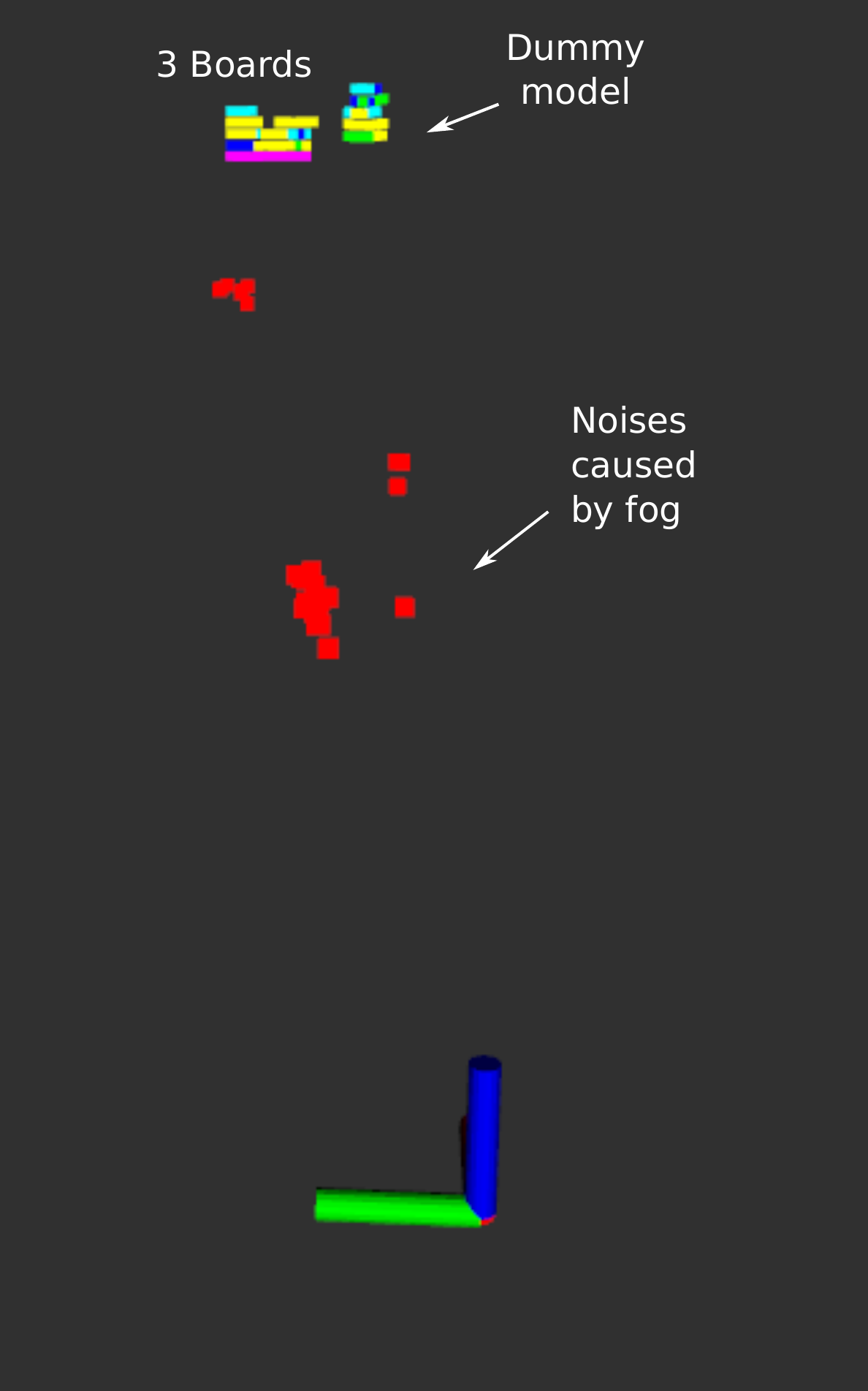}
}
\subfigure[LiDAR measures of targets (Two traffic signs, R=15m,V=15m)]{
\includegraphics[width = 0.22\textwidth, height=3.5cm]{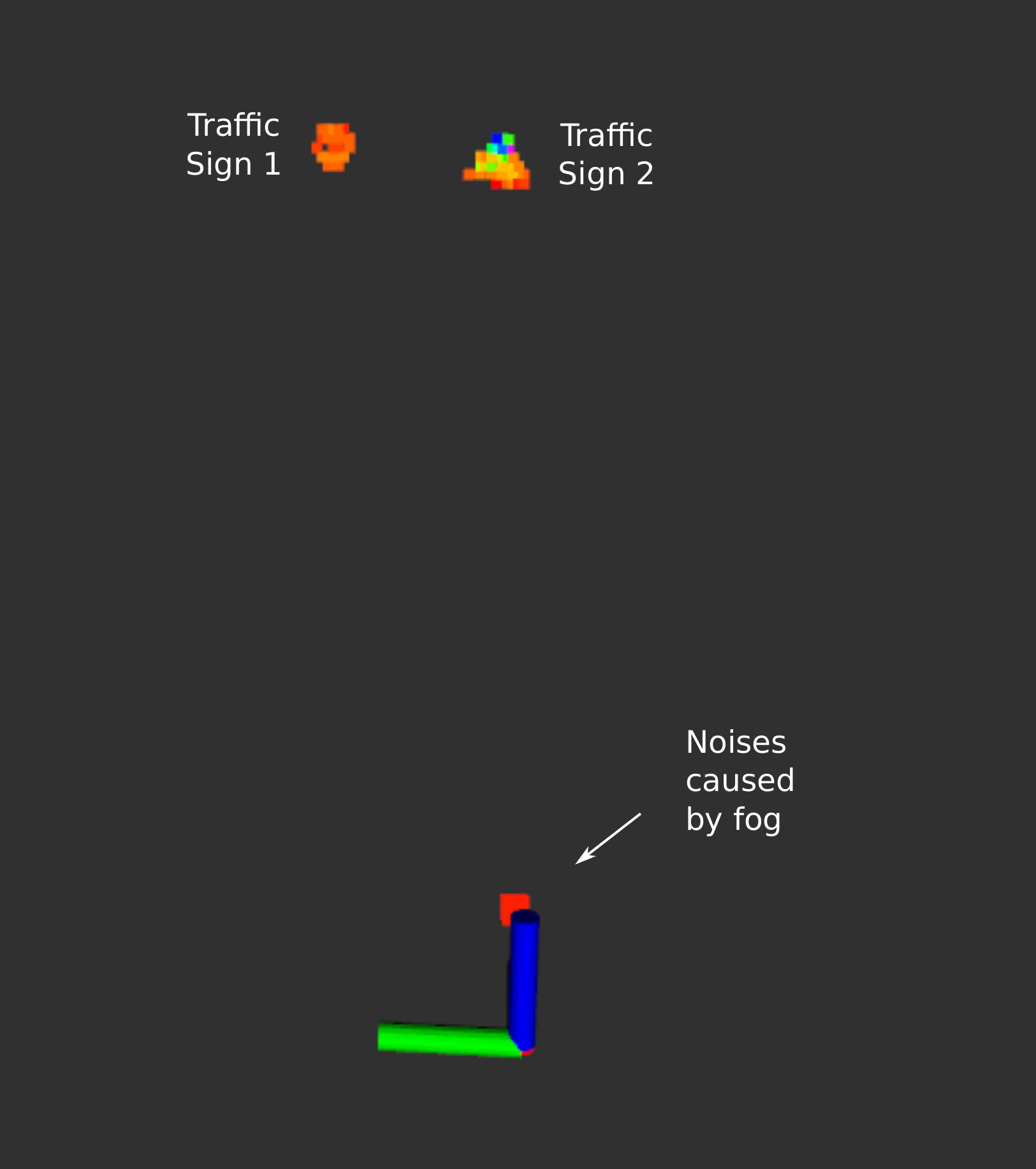}
}
\subfigure[LiDAR measures of targets (Car, R=10m, V=20m)]{
\includegraphics[width = 0.22\textwidth, height=3.5cm]{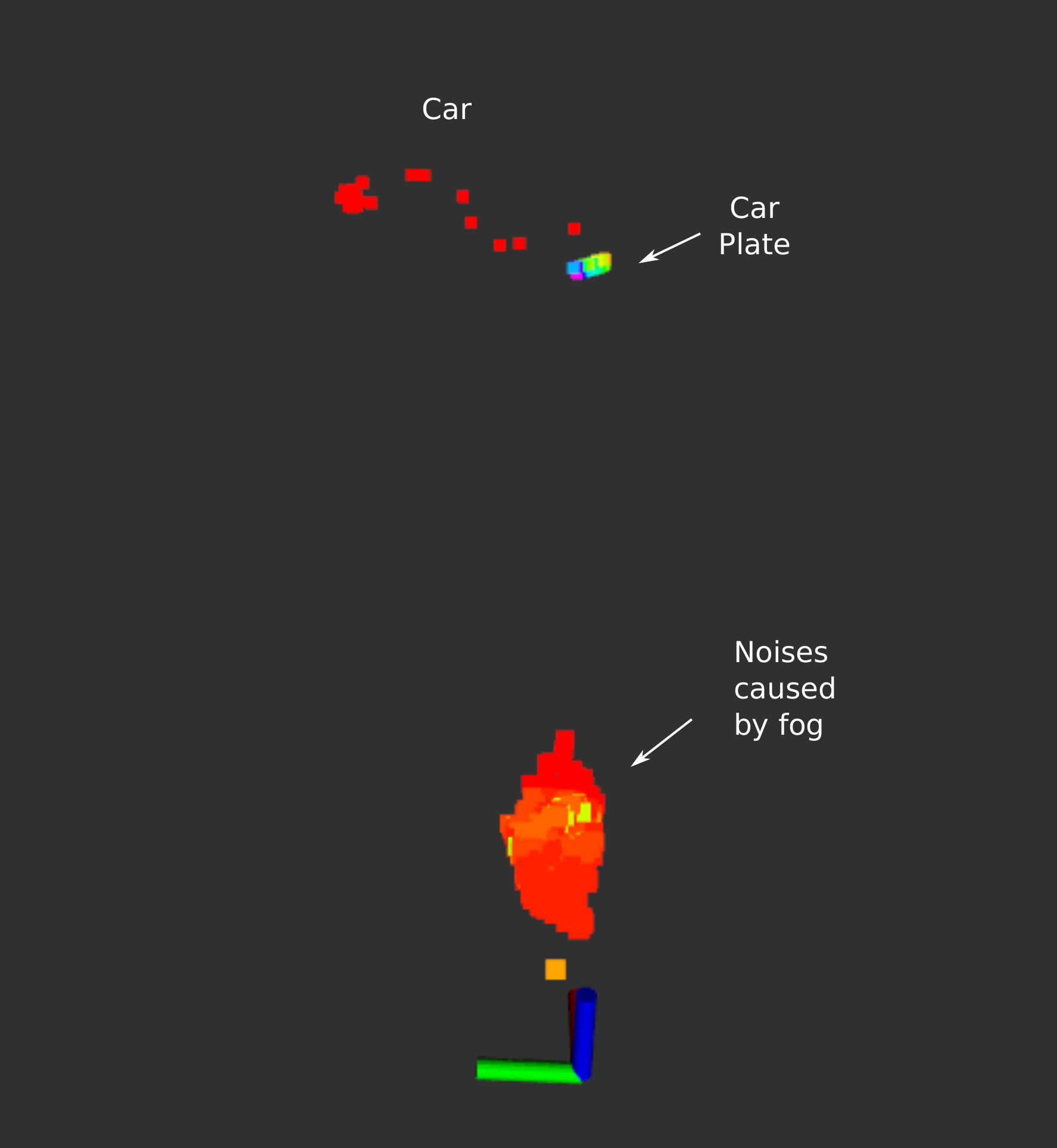}
}
\subfigure[LiDAR measures of targets (Car, R=10m, V=40m)]{
\includegraphics[width = 0.22\textwidth, height=3.5cm]{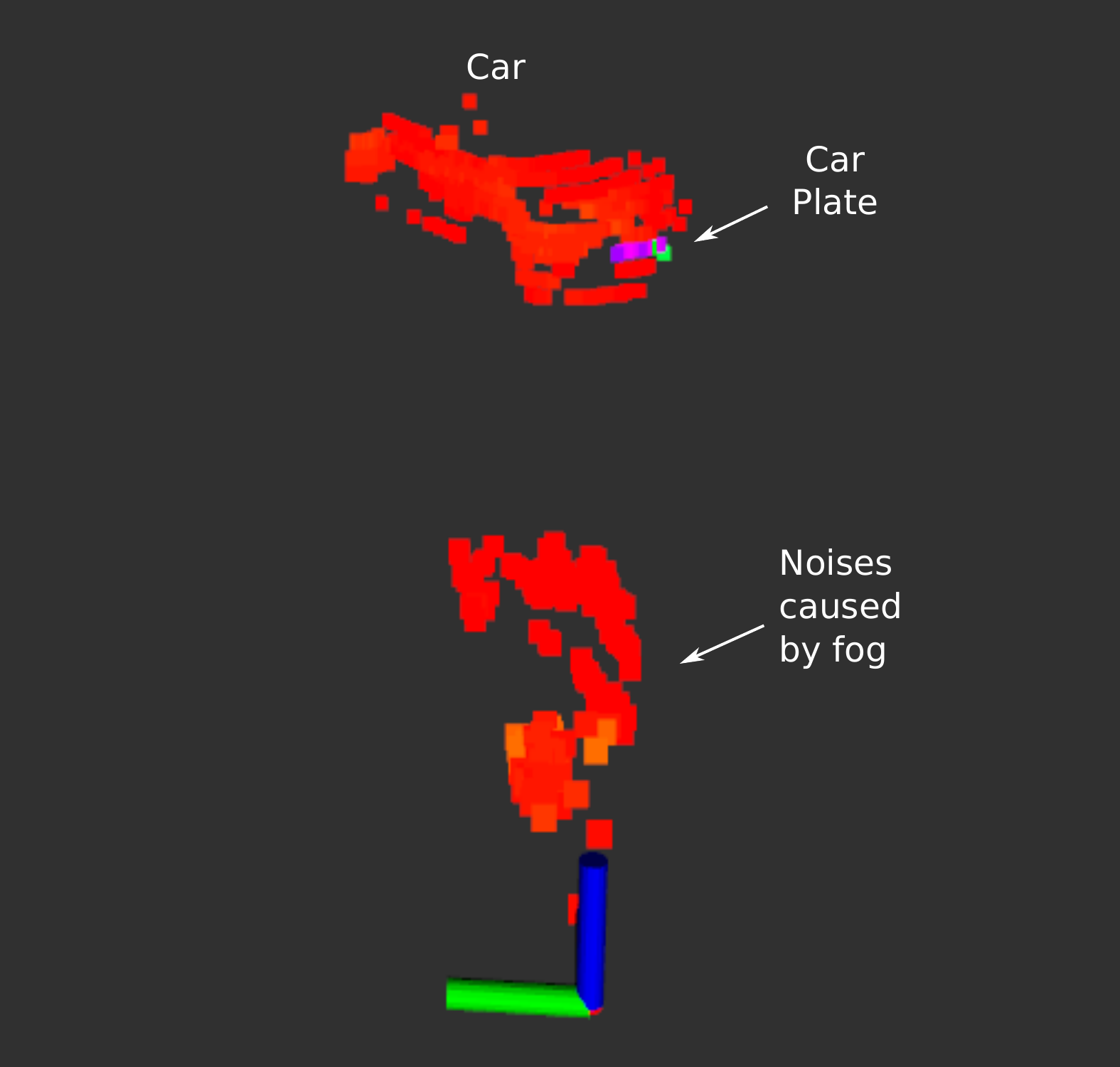}
}
\subfigure[LiDARmeasures of targets (Car, R=10m, V=80m)]{
\includegraphics[width = 0.22\textwidth, height=3.5cm]{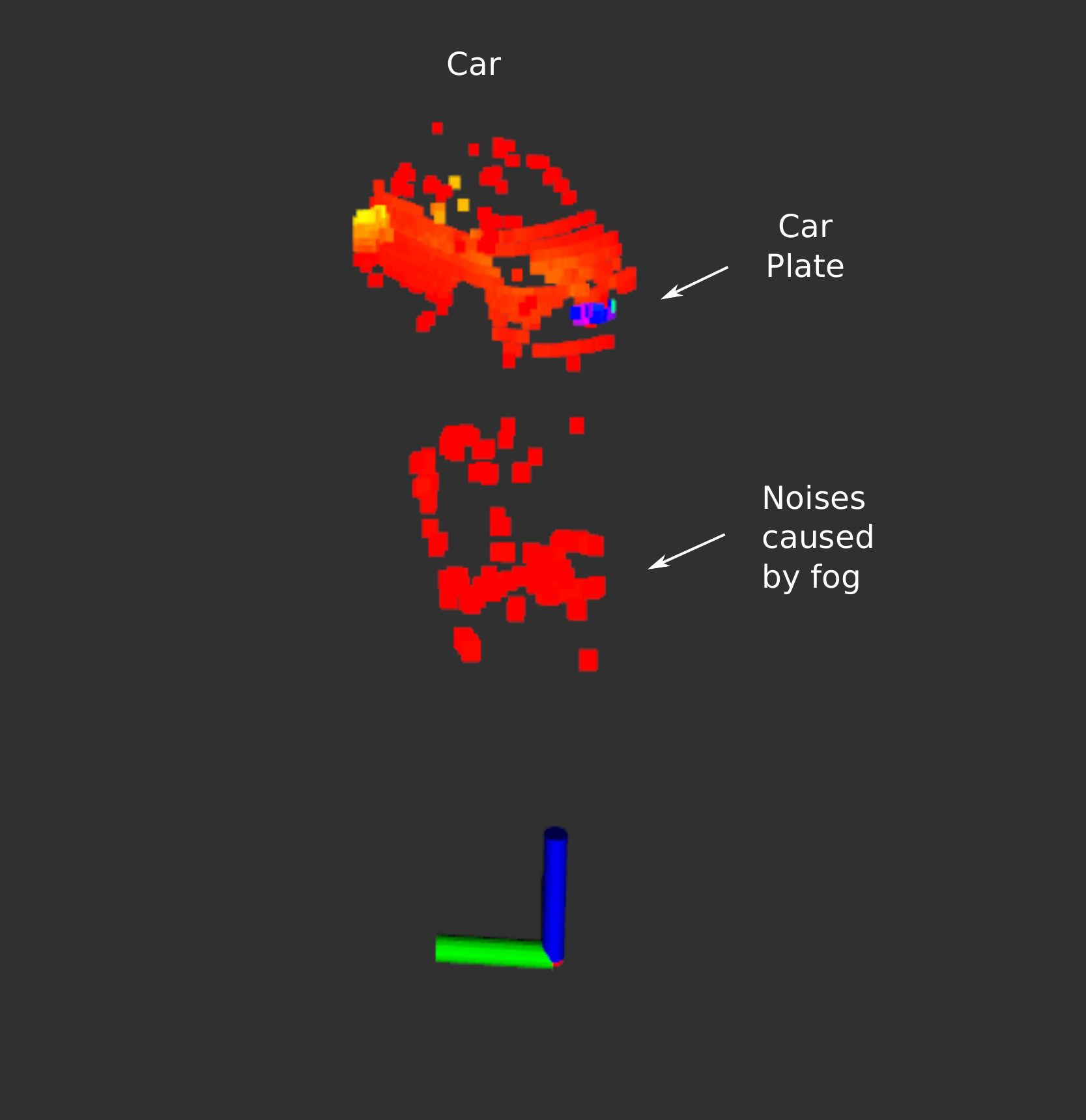}
}
\caption{Examples of LiDAR recordings in several scenarios. Color encodes the reflectivity. The axis represents the LiDAR.}
\label{fig::lidar_recording}
\end{figure*}

\subsection{LiDAR Recordings}
\begin{table}[b]
  \centering
  \begin{tabular}{m{1.8cm}|m{1.5cm}|m{4.5cm}@{}m{0pt}@{}}
    \toprule
    Weather Condition & Targets & Distance Target-LiDAR\\\hline
    \multirow{6}{*}{\tabincell{l}{\textbf{Fog dissipation:}\\meteorological \\ 10m visibility to\\ clear condition}} & Dummy model, three boards &10 times: 5-25m(every 2.5m),27meter\\\cline{2-3}
    & Car &4 times: 10-25m, every 5m&\\[0.7em]\cline{2-3}
              & Two traffic signs &5 times: 10-25m every 5m, 22.5m\\\cline{2-3}
    & None & Background ground truth&\\[0.7em]
    \bottomrule
  \end{tabular}
  \caption{All the tested scenarios}
\label{tab::exps}
\end{table}


In our tests, one or several targets were put at varying distances (from 5m to 30m) in front of the LiDAR. All the tested scenarios are summarized in Tab. \ref{tab::exps}. In each test, a ground truth LiDAR data was logged at the beginning without fog. Then, we started to generate the artificial fog controlled by visibility sensors. The LiDAR measurements, which contain \textit{range, azimuth angle, ring number, reflectivity, and timestamp}, were recorded until the targets were fully and stably detected. The logged LiDAR data was synchronized with meteorological visibility data as well.  

For each test, we manually extract a region of interest (ROI) of lasers hitting on the targets. For Velodyne UltraPuck, every transmitted laser can be indexed by a ring number (between 0 to 31) and an azimuth angle between 0 to 360 degrees, encoded by 0 -- 1800 when operating at 10HZ. The ring numbers and azimuth angles of the lasers hitting on the targets are manually extracted and saved as laser ROIs. Only LiDAR measures within the ROIs are retained for further analysis. For each test as in Tab. \ref{tab::exps}, all the recorded data can be represented as: 
\begin{equation}
  r_{i,j}(t), \beta_{i,j}(t), V(t), t | (i,j) \in ROI, t \in [t_0, t_1]
\end{equation}
where $i,j$ are the laser index comprising ring number and azimuth angle. $r_{i,j}(t)$, $\beta_{i,j}(t)$ and $V(t)$ are the range, reflectivity, and visibility measures at time $t$, respectively. $t_0$ and $t_1$ are the start time and end time of this test. 

\section{Analysis of Experimental Results}\label{sec::analyze}
\subsection{Modeling ranging process impacted by fog}
According the power model in Eq. \ref{eq::lidar}, the received laser power from the target is mainly decided by the distance $R$, surface reflectivity $\beta$ of the target, and the extinction coefficient $\alpha$ of the transmission medium. Since the signal processing unit is a blackbox embedded inside the sensor, we exclude this factor from consideration. The ranging process of a ToF LiDAR can be approximated as:
\begin{equation}
  r(t) \sim f(\alpha(t), \beta(t)|R, \beta )
\end{equation}
where $\beta(t)$ and $\alpha(t)$ are the target's reflectivity and the extinction coefficient during the fog test. $\beta$ is the target's reflectivity in clear condition. $f(\cdot)$ is a mapping function that outputs ranging results with regard to several variables such as $R, \beta, \alpha$.

In our paper, we assume the fog density has a even distribution through the ranging space. Under this homogeneity assumption of fog, the extinction coefficient $\alpha$ can be characterized by the meteorological visibility measurement $V$. Meanwhile, as introduced in Sec. \ref{sec::influence}, surface reflectivity $\beta$ is also influenced by fog, which is measured by visibility $V$. Therefore, during our tests, the fog impacts on ranging measures can be modeled as: 

\begin{equation}
  \begin{split}
  r(t) &\sim f(\alpha(t), \beta(t)|R,\beta) \\
  &\sim f(V(t)|R,\beta)
\end{split}
\label{eq::model}
\end{equation}
Where $f(\cdot)$ denotes a specific ranging process under such situation. Eq. \ref{eq::model} qualitatively illustrates that, during our test, the LiDAR performance is comprehensive influenced by three factors: the distance $R$, the target's reflectivity $\beta$, and the severity of fog $V(t)$. Due to the ignorance of the internal signal processing unit, analytical form of $f(\cdot)$ can hardly be obtained.

\subsection{Qualitative analysis}
Based on the models in Eq. \ref{eq::lidar} and Eq. \ref{eq::model}, we can infer the following LiDAR characteristics under fog:

\begin{figure*}[t]
  \centering
\subfigure[]{
\includegraphics[width = 0.31\textwidth]{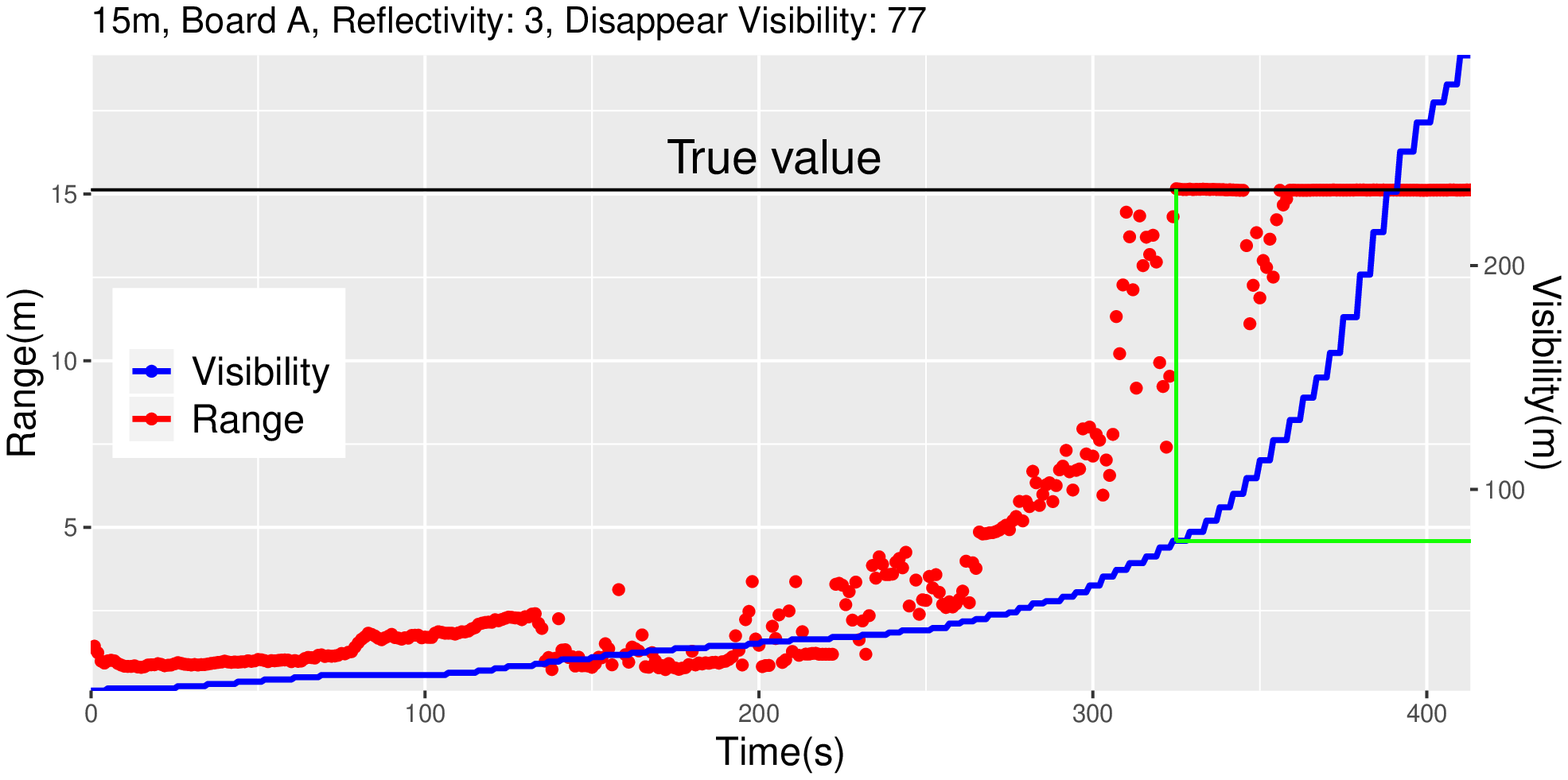}
}
\subfigure[]{
\includegraphics[width = 0.31\textwidth]{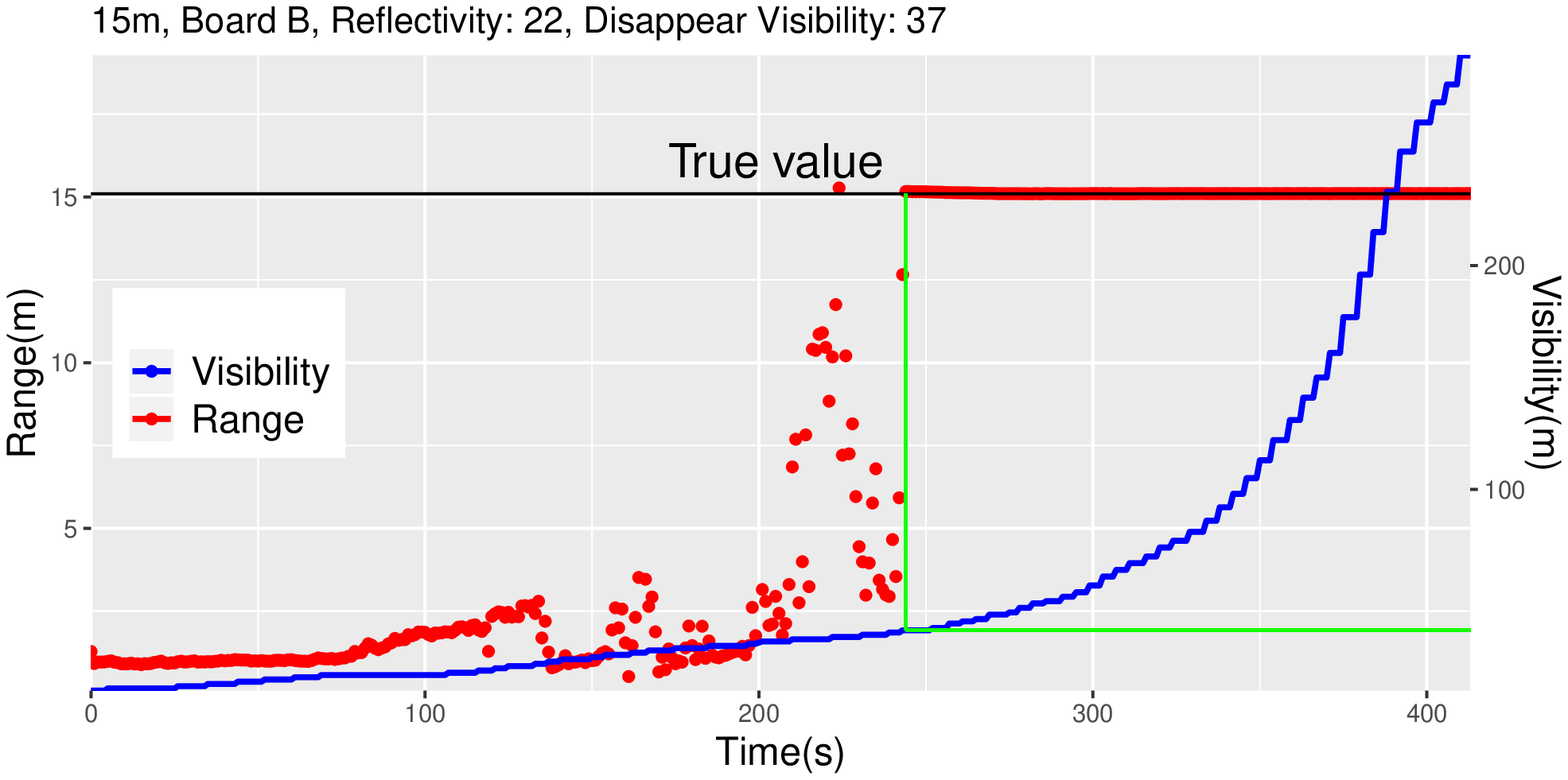}
}
\subfigure[]{
\includegraphics[width = 0.31\textwidth]{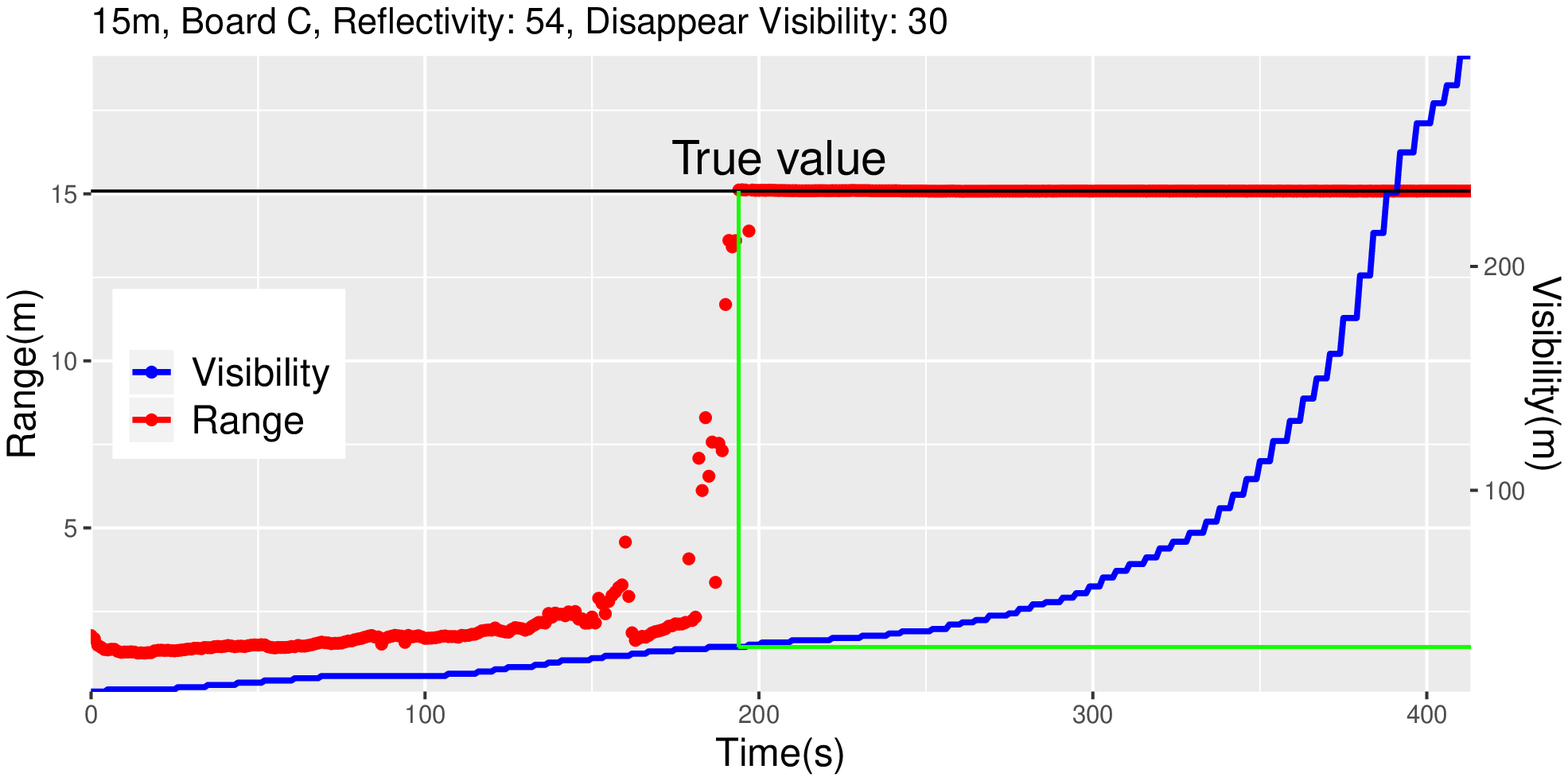}
}
\subfigure[]{
\includegraphics[width = 0.31\textwidth]{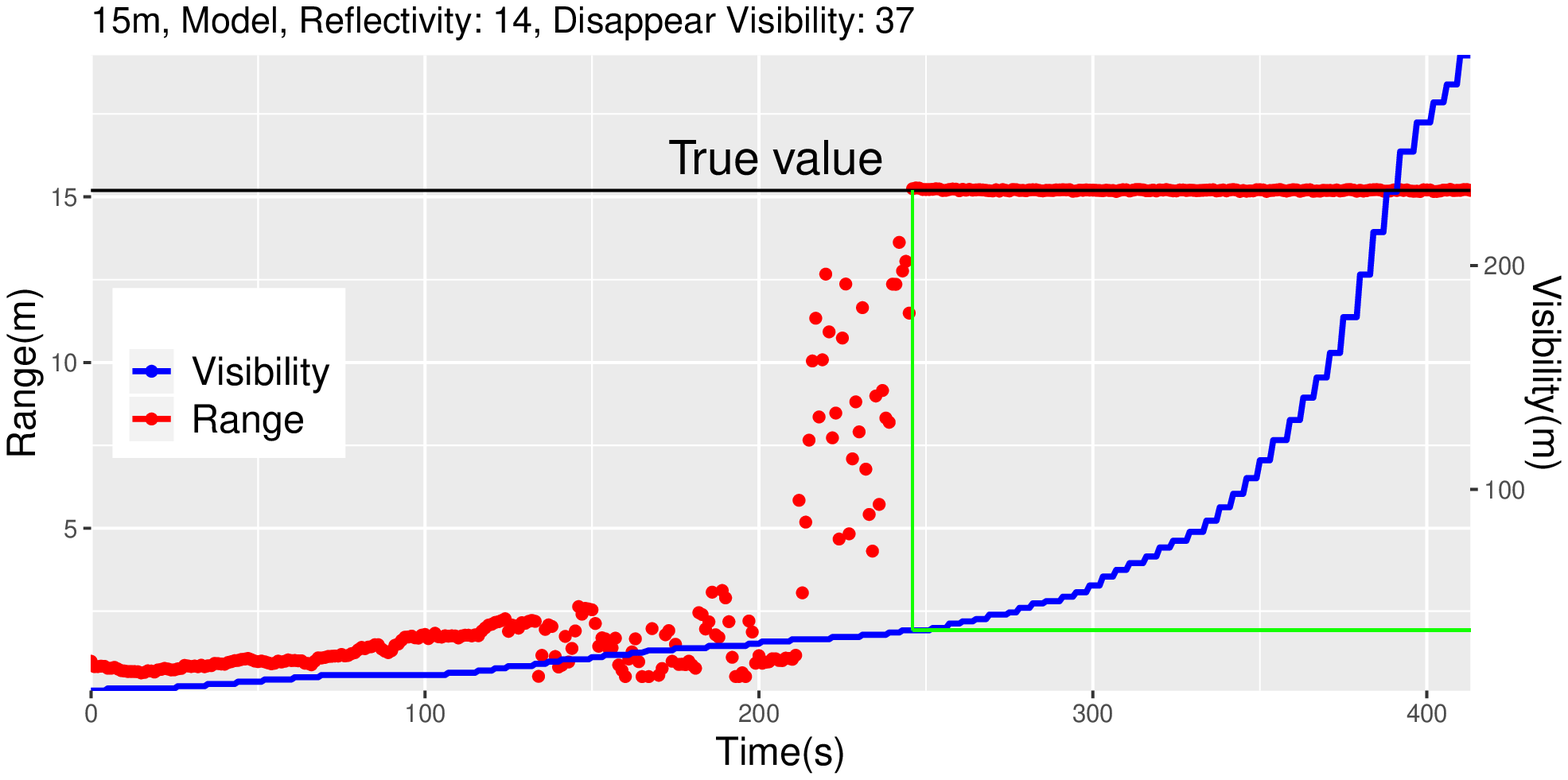}
}
\subfigure[]{
\includegraphics[width = 0.31\textwidth]{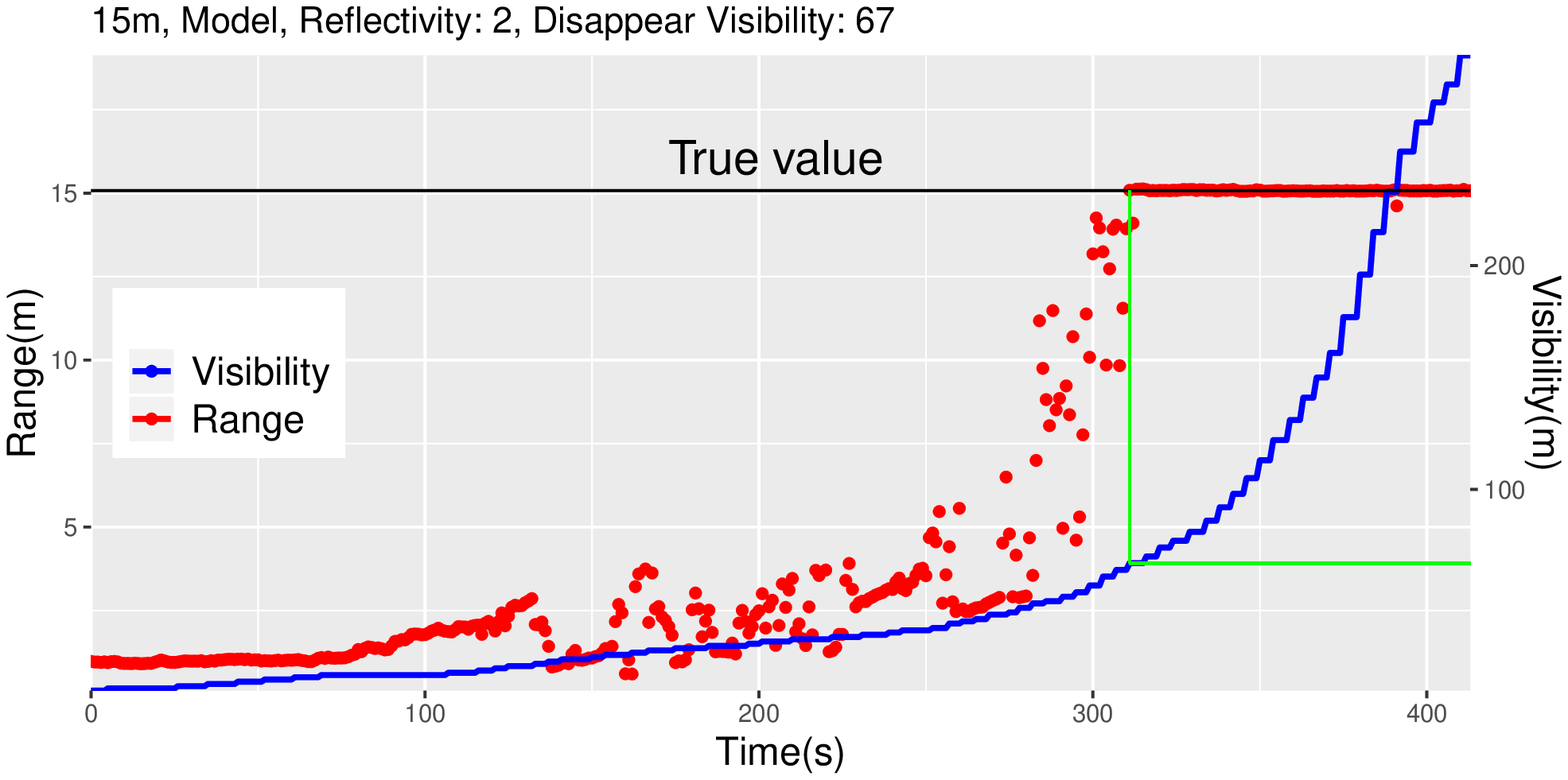}
}
\subfigure[]{
\includegraphics[width = 0.31\textwidth]{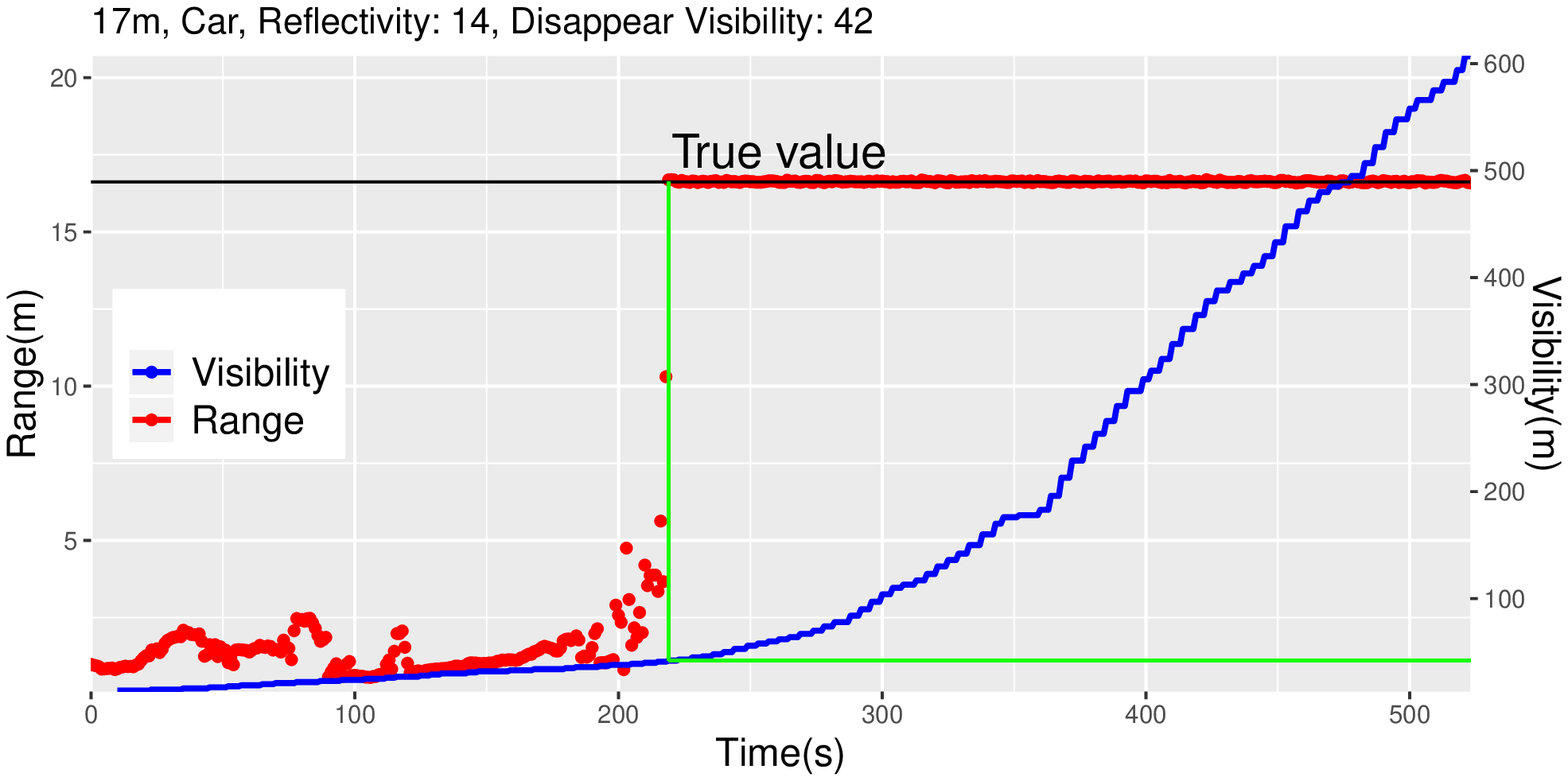}
}
\subfigure[]{
\includegraphics[width = 0.31\textwidth]{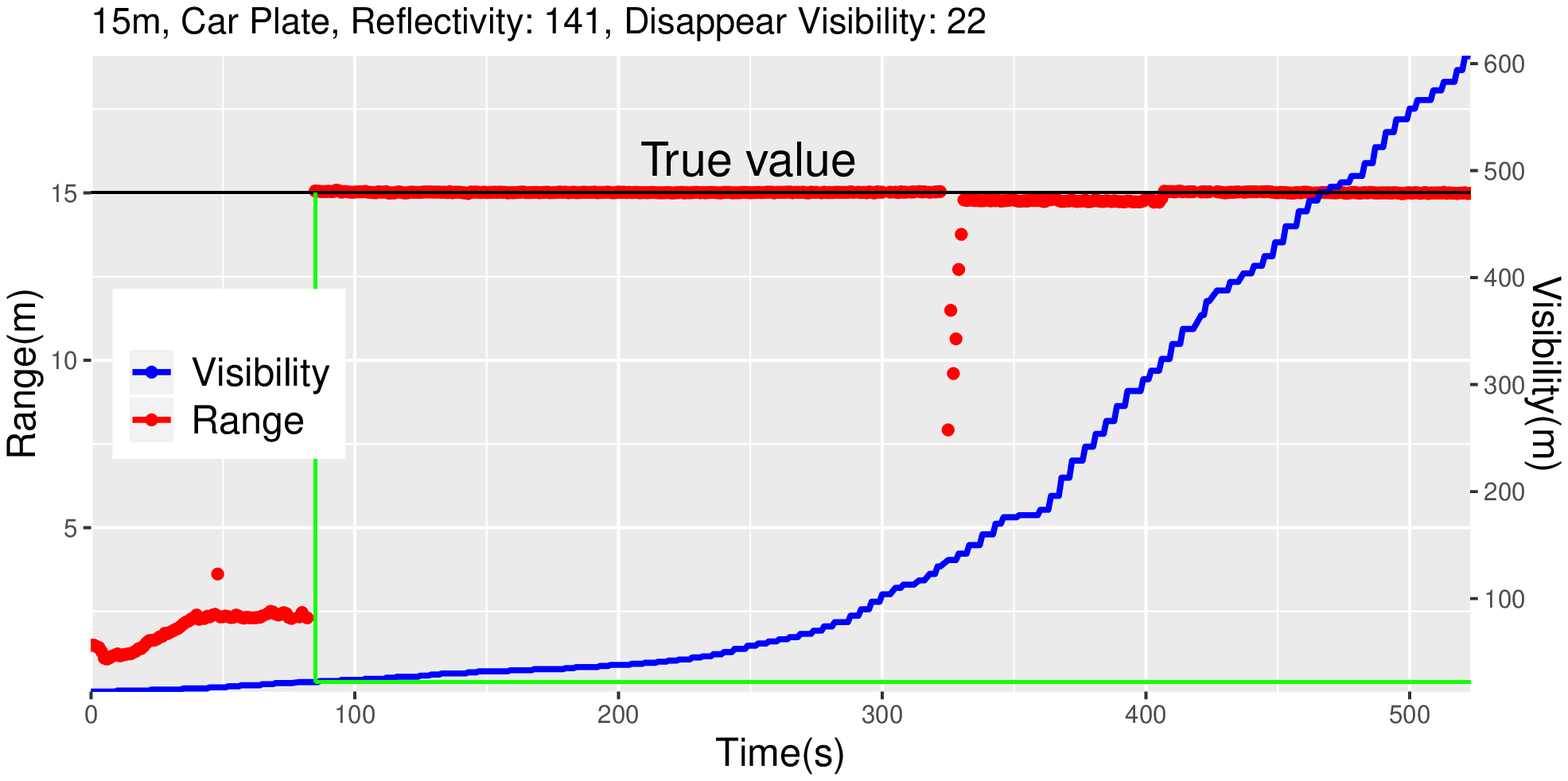}
}
\subfigure[]{
\includegraphics[width = 0.31\textwidth]{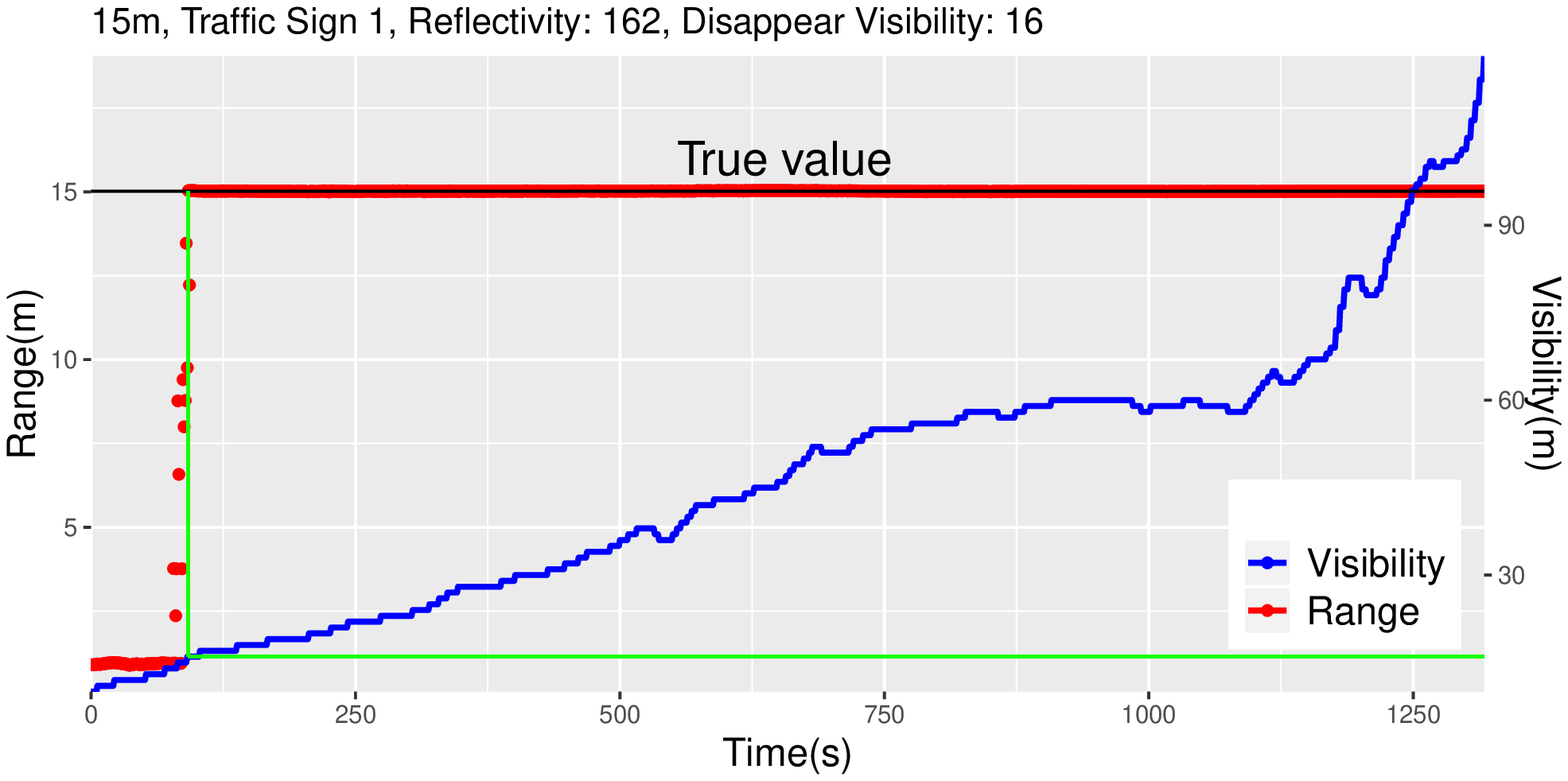}
}
\subfigure[]{
\includegraphics[width = 0.31\textwidth]{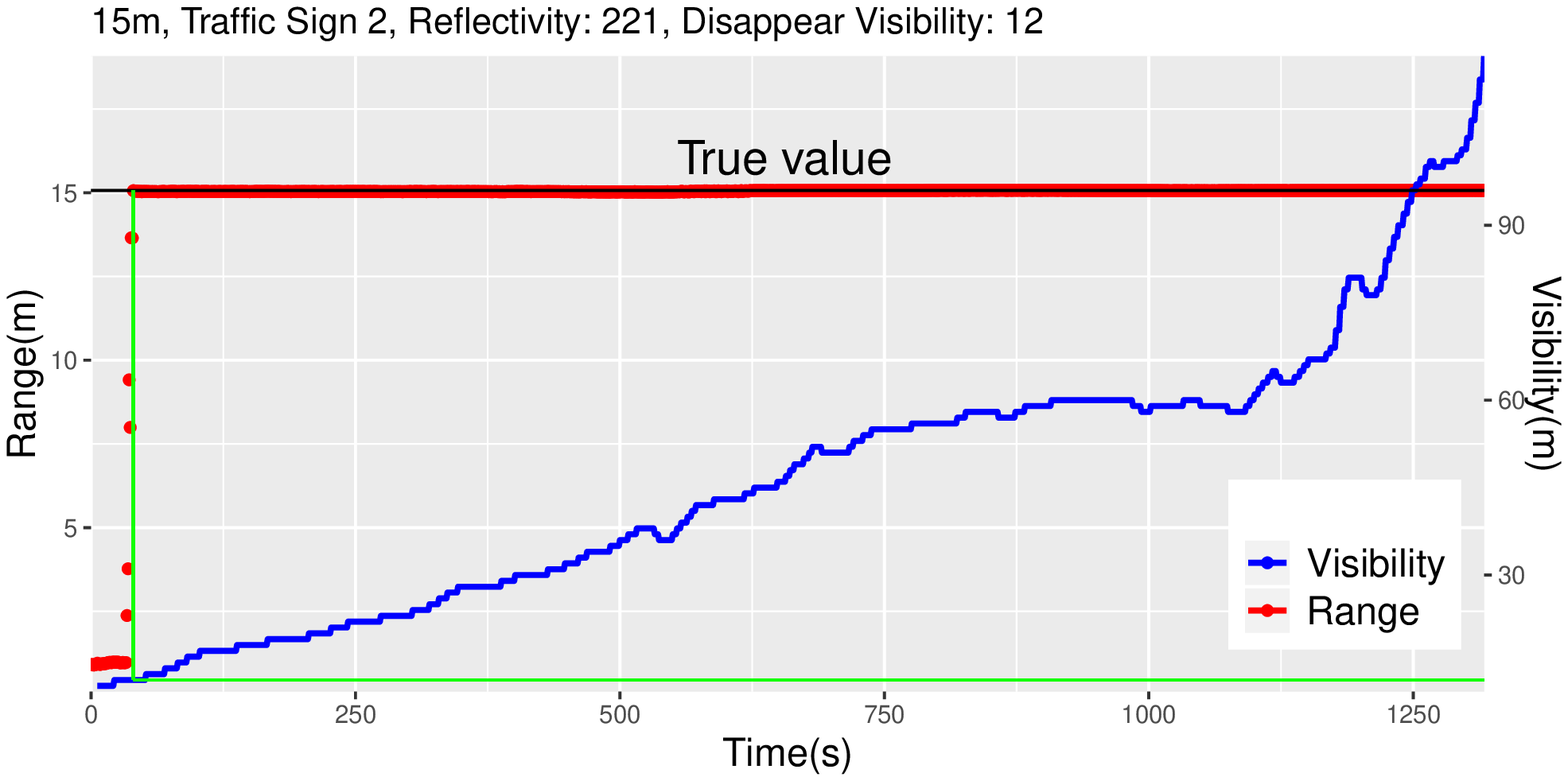}
}
\caption{Average range measures $\bar{r}_{i,j}(\lfloor t \rfloor)$ with meteorological visibility $V(t)$ for randomly selected individual lasers of various targets at 15m. Disappear visibilities are marked as the intersections between the green lines and the axis of "Visibility".}
\label{fig::process}
\end{figure*}

\begin{enumerate}
\item For a target at given distance $R$, visibility $V$ is \textit{proportional to} the ranging capability: the higher the visibility $V(t)$ at time $t$, the less the range error $|r(t) - R|$.   
\item For a given visibility $V$ and given distance $R$, surface reflectivity $\beta$ is \textit{proportional to} the ranging capability: the higher the target's surface reflectivity, the less the range error $|r(t) - R|$.
\item For a target under a certain visibility fog condition, distance $R$ is \textit{reverse proportional to} the ranging capability: the bigger the $R$, the bigger the range error $|r(t) - R|$.
\end{enumerate} 

Those three characteristics can be verified by the tests summarized in Tab. \ref{tab::exps}. Fig. \ref{fig::lidar_recording} visualizes the LiDAR measures of several objects under various visibilities and distances. Fig. \ref{fig::lidar_recording} (a) and (b) show the difference of LiDAR outputs between clear and foggy environments. The clutter points in (b) are the ranging noises caused by fog, and part of the boards are not detected comparing with (a). Fig. \ref{fig::lidar_recording} (c) and (d) show the range measures for the three calibrated boards and dummy model at 15m with visibility of 40m and 80m respectively. In Fig .\ref{fig::lidar_recording} (c), board A ($5\%$ reflectivity, on the left) and the lower part of the model are barely visible, while all targets are detected in (d). Fig .\ref{fig::lidar_recording} (e) and (f) demonstrate similar phenomenon when the targets are at 20m. The comparison between (c)-(d) and (e)-(f) verify the relation between ranging capability and visibility. From Fig .\ref{fig::lidar_recording} (c) to (f), we also observe that the objects with higher reflectivity demonstrates better robustness under fog. The board C of strongest reflectivity ($90\%$, in the middle) is detected before the other two boards, as comparing (e) with (f). Comparing (f) with (d), we can find that, under the same visibility, board A and the lower part of the model do not appear in (f). However, these parts are detected in (d) when the distance is smaller. This reveals the third property summarized above, which can also be found from comparing (c) with (e).

Fig. \ref{fig::lidar_recording} (g) shows the testing results of two traffic signs at 15m. Since the reflectivity of two traffic signs are much higher than the others (as in Fig. \ref{fig::method} (e)), they are fully detected even when $V=15m$ -- much less than the test of three boards as shown in (d) when $V=80m$. As the tests of car demonstrated in Fig. \ref{fig::lidar_recording} (h)~(j) no surprise to observe that the car's plate appears earlier than other parts. Because the tire and the window parts of the car have the lowest reflectivities, those parts are still invisible for the LiDAR when the other parts are detected, as shown in Fig. \ref{fig::lidar_recording} (j). All the three qualitative properties summarized above can be verified in the test examples.

\subsection{Quantitative analysis}
\begin{figure*}[t]
  \centering
\subfigure[The disappear visibilities for three boards]{
\includegraphics[width = 0.46\textwidth]{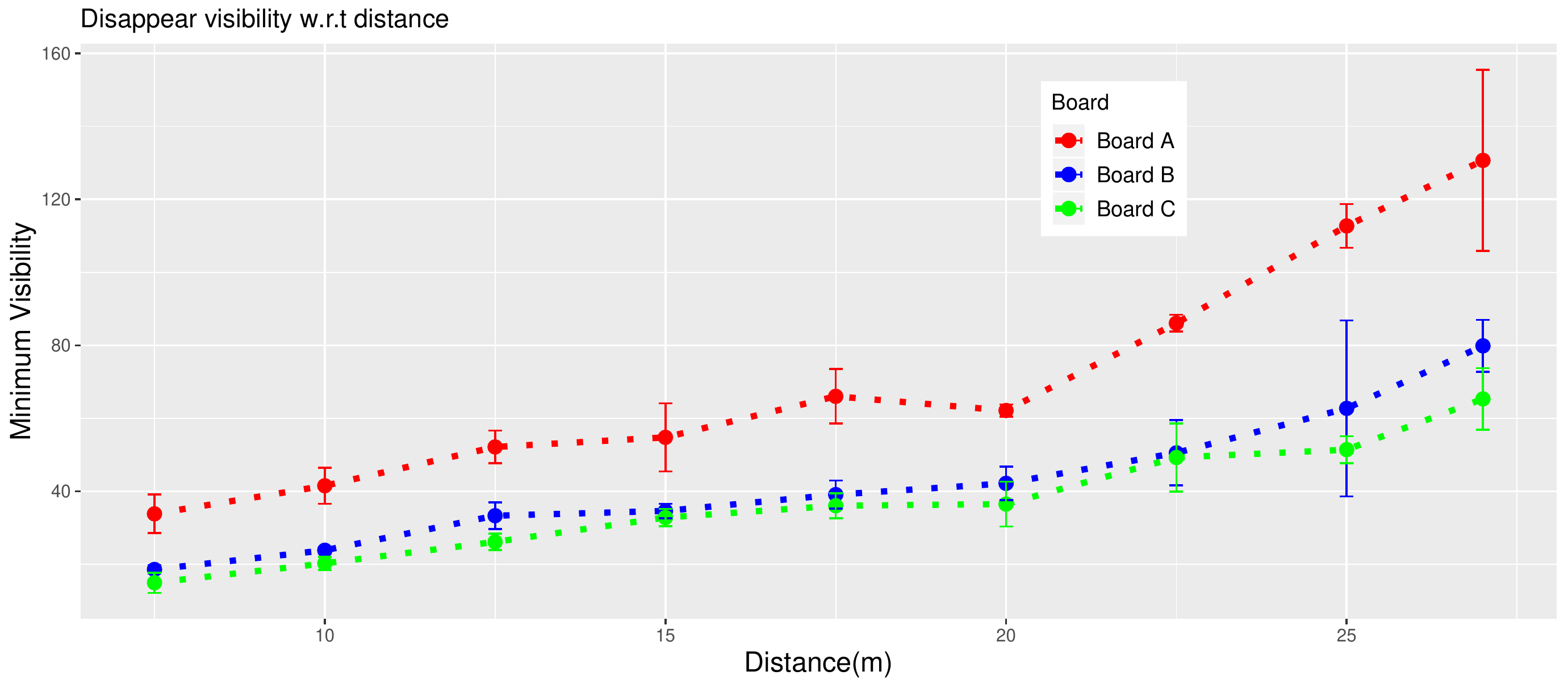}
}
\subfigure[The disappear visibilities for the two traffic signs]{
\includegraphics[width = 0.46\textwidth]{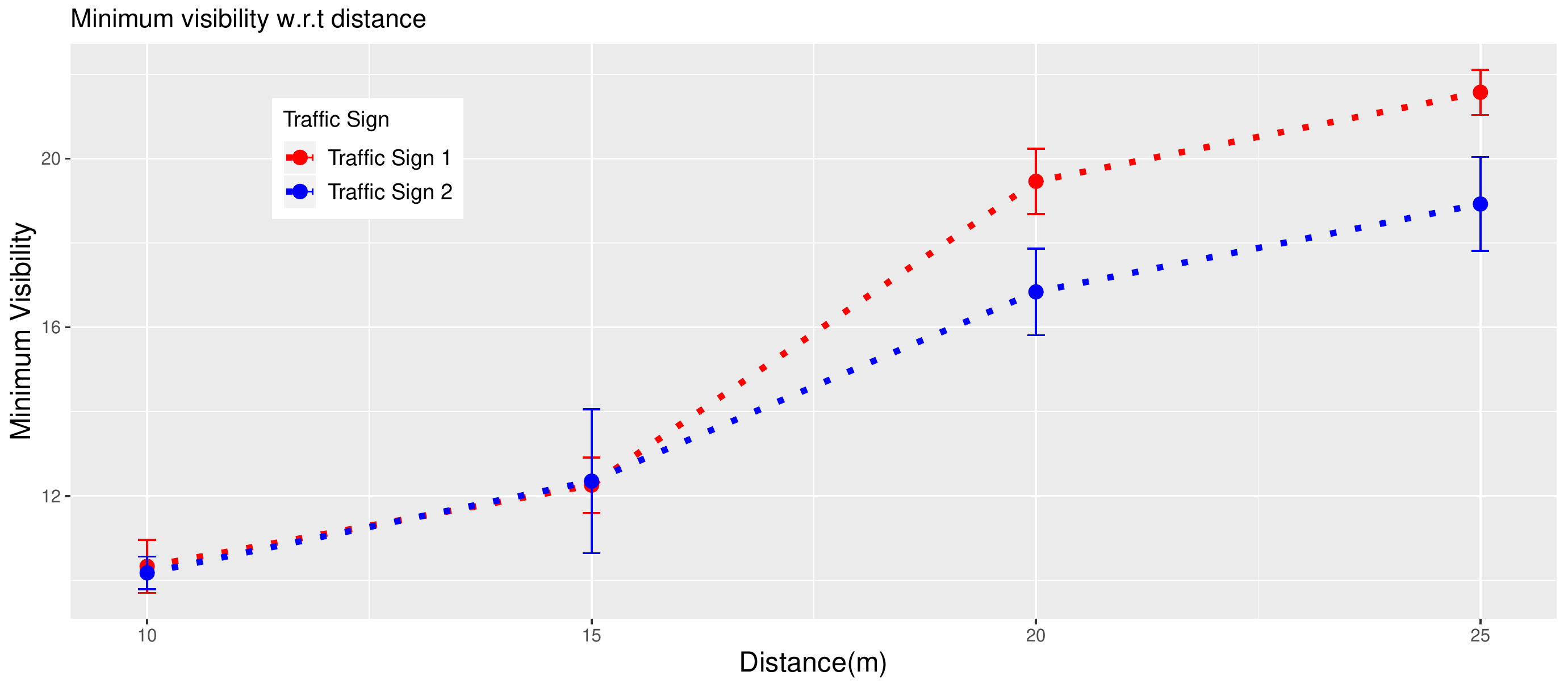}
}
\subfigure[The disappear visibilities for the model (divided by upper part and lower part)]{
\includegraphics[width = 0.46\textwidth]{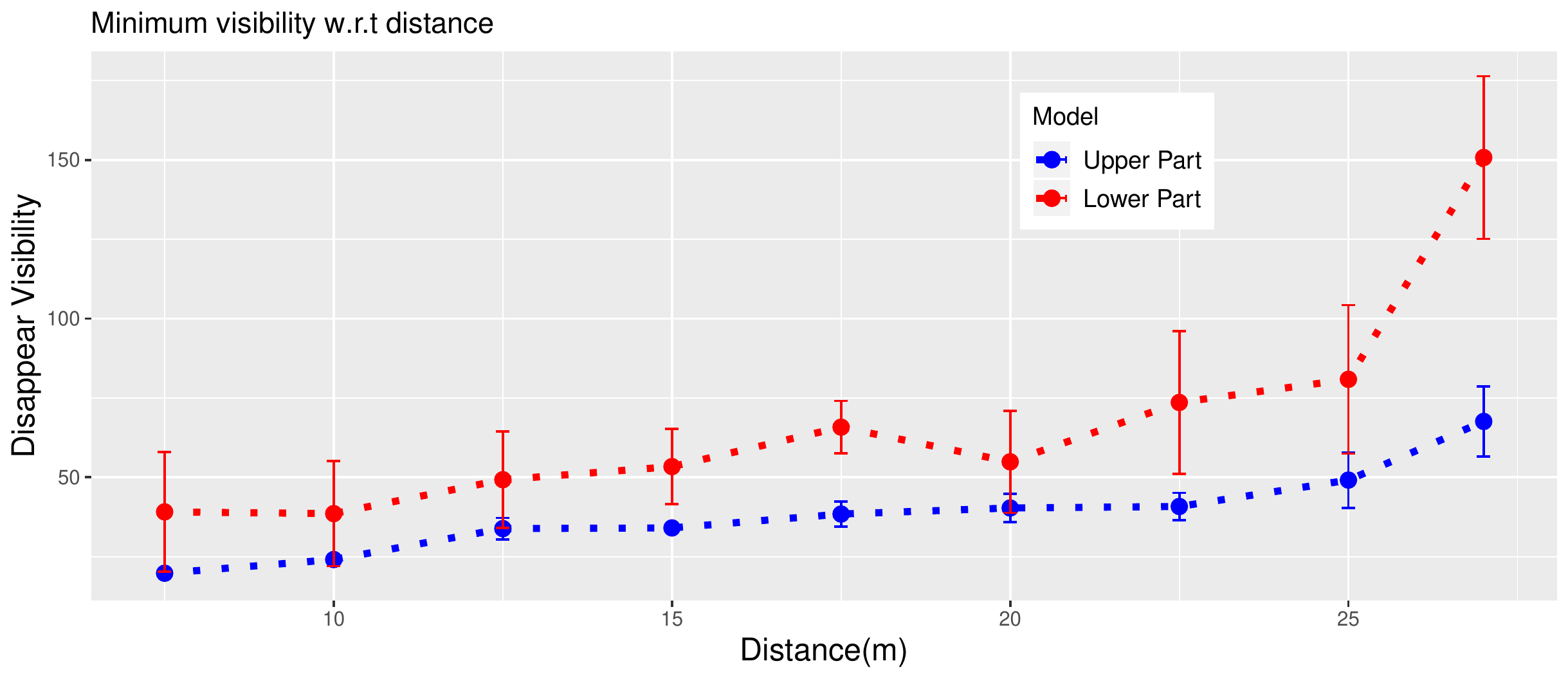}
}
\subfigure[The disappear visibilities for the car (divided by plate, strong reflection and weak reflection parts]{
\includegraphics[width = 0.46\textwidth]{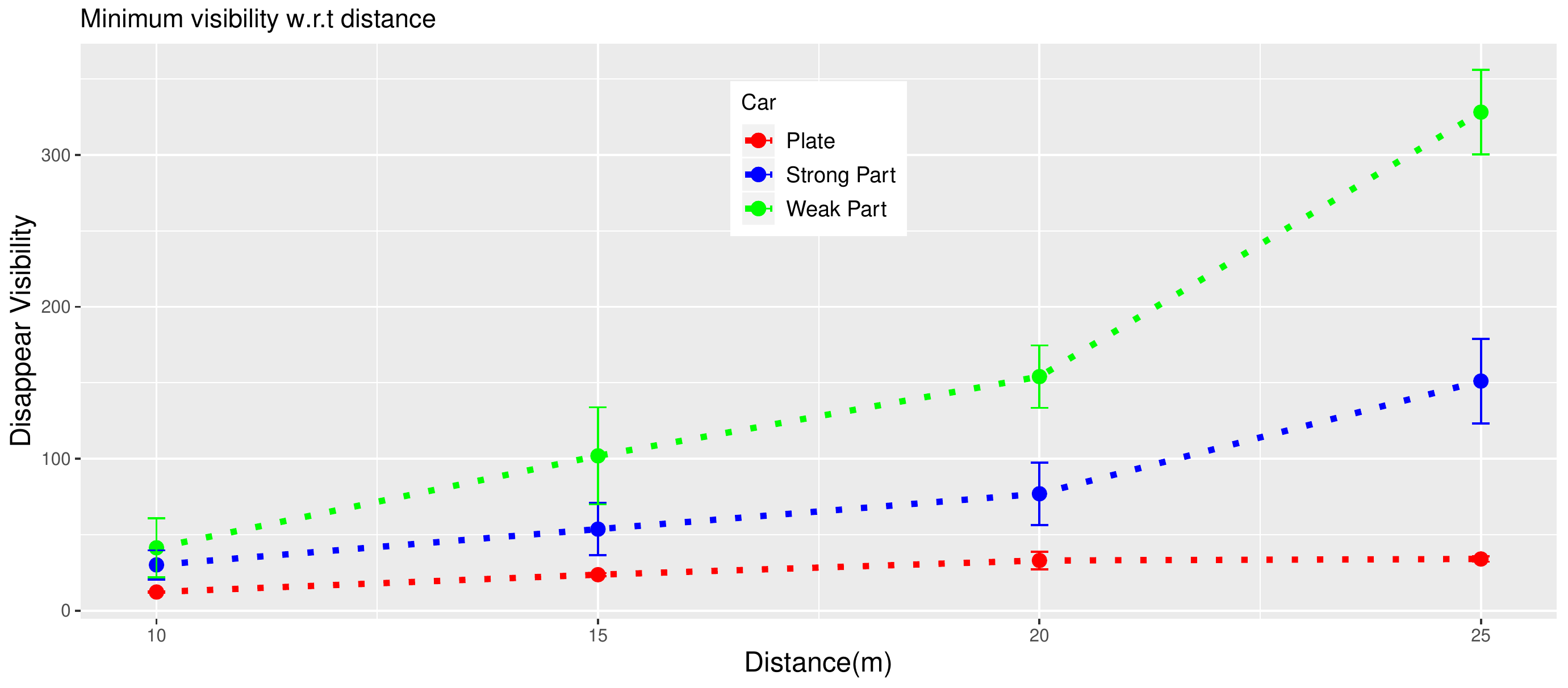}
}
\caption{The disappear visibilities for different objects with regard to various distances}
\label{fig::min_vis}
\end{figure*}

The above qualitative analysis gives a general picture of Velodyne UltraPuck's performance with regard to distance, targets and fog density. In this section, we quantitatively evaluate the behavior of Velodyne UltraPuck in foggy environment.

\subsubsection{Individual ranging process} Being synchronized with meteorological visibility data, the ranging process $r_{i,j}(t)$ of each individual lasers within ROI is able to be visualized. Since the LiDAR is running at 10HZ while the visibility data is 1HZ, we utilize the average range measures $\bar{r}_{i,j}(\lfloor t \rfloor)$ during every second:
\begin{equation}
  \bar{r}_{i,j}(\lfloor t \rfloor) \triangleq \frac{1}{10}\sum_{\lfloor t \rfloor}^{\lfloor t+1 \rfloor}{r_{i,j}(t)}, \lfloor t \rfloor \in [0,1,2,...]
  \label{eq::average}
\end{equation}
Fig. \ref{fig::process} (a) - (i) visualize the $\bar{r}_{i,j}(\lfloor t \rfloor)$ of several randomly selected lasers within ROI, their true ranges without fog and the synchronized visibility measures. Fig. \ref{fig::process} (a) plots the $\bar{r}_{i,j}(\lfloor t \rfloor)$ (red) and $V(t)$ (blue) for a certain laser hitting on Board A (measured ground truth reflectivity 3 by LiDAR) at 15m. In the beginning, due to low visibility, the measured range starts from false values much smaller than the true value (15m). Then, along with the fog dissipation, $\bar{r}_{i,j}(\lfloor t \rfloor)$ gradually increases until reaches the true value. After reaching true value, although sometimes $\bar{r}_{i,j}(\lfloor t \rfloor)$ deviates from the ground truth, it is generally stable. The range measure's trend of increasingly close to true value with regard to the visibility augment is clearly observed. Similar tendency can be discovered from (b) to (i), which are samples of LiDAR measurements for the other targets. 

\subsubsection{Disappear visibility}
Apart from quantitatively demonstrating the relation between the range measures and meteorological visibility, through Fig. \ref{fig::process} (a) - (i), we can find the time when the LiDAR measures start to be true values. By associating this time with the recorded meteorological visibility, we can define a \textit{disappear visibility} representing the ranging capability:
\begin{myDef}
  \label{definition}
  Given a certain distance $R$ and target's surface reflectivity $\beta$, \textit{disappear visibility}  is the minimum visibility that allows the correspondent ranging process return the true distance measure. For an individual laser $[i,j]$ within ROI in our tests, its disappear visibility is:
\begin{equation*}
\begin{aligned}
 V_{i,j}^{dis}|R,\beta\quad\triangleq \quad & \underset{V}{\text{minimize}}
& & V_{i,j}(t)|R,\beta \\
& \text{subject to}
& & |r_{i,j}(t)-R|<\sigma \\
& \text{where}
& & r_{i,j}(t) \sim f(V(t)|R,\beta)
\end{aligned}
\label{eq::def_vis}
\end{equation*}
\end{myDef}
Where $\sigma$ is a small threshold that decides whether the measured range equals to the truth or not. In Fig. \ref{fig::process} (a), the green lines point out the disappear visibility (77m) for a scanned point on Board A (relectivity 3) at 15m distance. The disappear visibilities for other tests are also shown in Fig. \ref{fig::process} (b) - (i).

$V^{dis}$ is an important indicator describing a LiDAR's measurability under fog environment.  As an important indicator describing the measurability of a LiDAR under fog environment, $V^{dis}$ has two-fold meanings or usages:
\begin{itemize}
\item For a given obstacle at a given distance, $V^{dis}$ can be used for benchmarking different types of LiDARs. A low $V^{dis}$ represents a good anti-interference capability within fog. 
\item For a given LiDAR, $V^{dis}$ points out its functionality under fog environment. Therefore, for a nature fog with measured visibility $V$, comparing with $V^{dis}$ can provide an evaluation of operational feasibility for a LiDAR based autonomous vehicle.
\end{itemize}

Fig. \ref{fig::min_vis} (a) - (d) visualize the $V^{dis}$ for all the tested objects. The 3 calibrated boards and 2 traffic signs can be assumed to have homogeneous surface of reflectivity, while for the car and dummy model, we group the measured reflectivities into similar clusters. The average reflectivities for each targets (at 15m) are: Board A: 2.75, Board B: 22.2, Board C: 45.54, Model Upper part: 17.7, Model lower part: 2.67, traffic sign 1: 169.9, traffic sign 2: 209.2, Car plate: 133.04, car strong part: 15.6, car weak part: 1.17.

From those experimental results, the disappear visibility is principally influenced by the surface reflectivity. In general, the stronger reflectivity, the lower disappear visibility. In Fig. \ref{fig::min_vis} (a), the order of disappear visibilities of Board A/B/C aligns with the order of their average reflectivities: $\beta_A<\beta_B<\beta_C, V^{dis}_A>V^{dis}_B>V^{dis}_C$. This effect is repeatedly verified by the average disappear visibilities for the model, car and traffic signs. As the traffic sign 2 has the highest reflectivity measures, it is not surprising to observe that it has the lowest disappear visibility.

\subsection{Modeling disappear visibility by machine learning}\label{sec::gpr}
After discussing the experimental results of LiDAR measures (in Fig. \ref{fig::lidar_recording}, \ref{fig::process}) and the disappear visibilities (in Fig. \ref{fig::min_vis}), we are interested to model the disappear visibility based on the recorded data. However, for a specific LiDAR, giving an analytical form of $V^{dis}$ for a certain target at a certain distance is too complicated to be achieved. Therefore, based on the recorded dataset in CEREMA's adverse weather facility, we propose a data-driven method to model $V^{dis}$ for the tested LiDAR.        

From Eq. \ref{eq::model} and the definition of $V^{dis}$, we can simply infer that $V^{dis}$ is influenced by the $R, \beta$:
\begin{equation}
  V^{dis} \sim g(R,\beta)
  \label{eq::gpvis}
\end{equation}
where $g(\cdot)$ is a function implying the relationship between $V^{dis}$ and $R,\beta$.

\begin{figure}[t]
  \centering
{
\includegraphics[width = 0.4\textwidth]{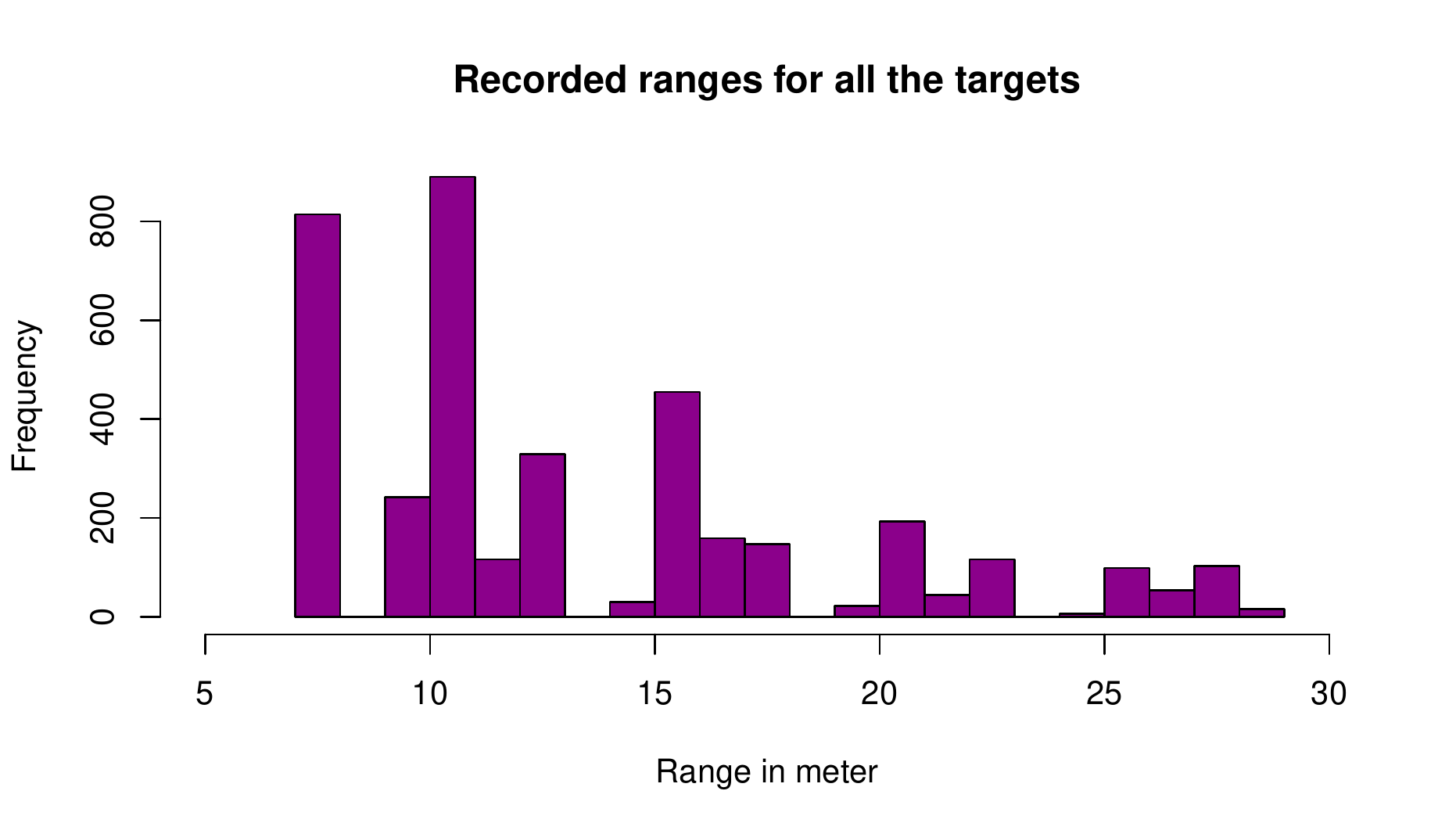}
}
{
\includegraphics[width = 0.4\textwidth]{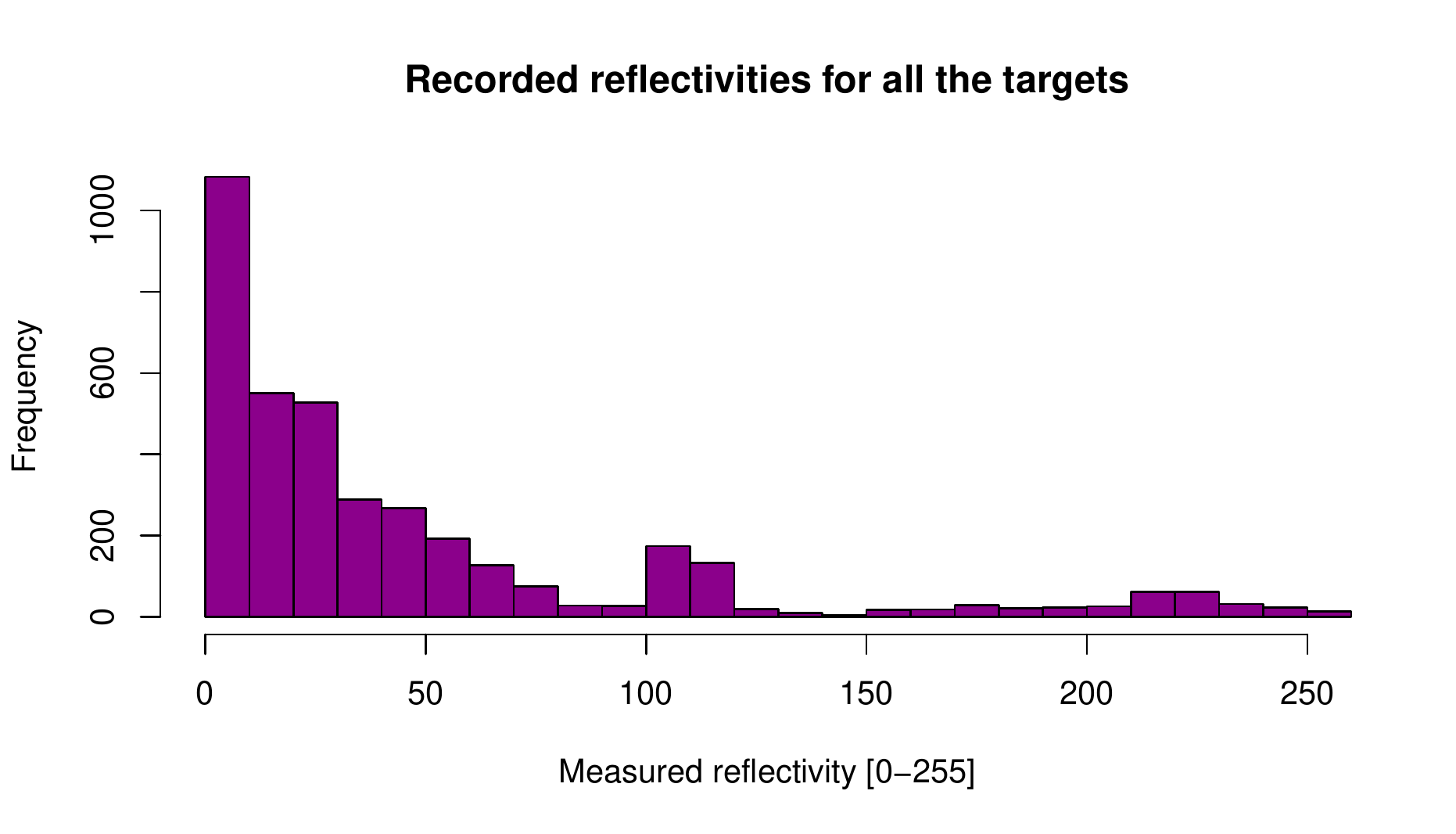}
}
\caption{The histograms of ranges (top) and reflectivities (bottom) for the recorded data}
\label{fig::hist}
\end{figure}

\subsubsection{Gaussian Process Regression (GPR)} \label{sec::gpm}
Gaussian process (GP) \cite{GPML} is a non-parametric machine learning tool that do not give an explicit function between the inputs and outputs. A Gaussian process is an infinite dimensional Gaussian distribution. A GP model is entirely defined by a mean function $m(\mathbf{x})$ and a covariance function $k(\mathbf{x},\mathbf{x}')$:
\begin{equation}
  f(\mathbf{x}) \sim \mathcal{GP}(m(\mathbf{x}), k(\mathbf{x},\mathbf{x}'))
\label{eq::gp_def}
\end{equation}
Usually we assume $m(\mathbf{x})=0$. One of the most popular usages of GP is regression. Having a training set $\mathcal{D}$ of $n$ observations, $\mathcal{D} = {(\mathbf{x}_i,y_i)| i = 1,...,n}$, where $\mathbf{x}$ denotes a $D$-dimensional input vector and $y$ denotes a scalar output, we assume the observations have additive i.i.d Gaussian noise with variance $\sigma_n^2$: $y = f(x)+\varepsilon$. We are interested in making inferences based on the relationship $f(\cdot)$ between inputs and outputs. Under the framework of GP, an inference for a test point $\mathbf{x}^*$ involves the computation of the mean $\bar{f}(\mathbf{x}^*) = \bar{f^*}$ and variance $\mathbb{V}[f^*]$:
\begin{equation}
  p(f_*|\mathbf{x},y, \mathbf{x}_*)  \sim \mathcal{N}(\bar{f_*}, \mathbb{V}[f_*])
  \end{equation}
\begin{equation}
  \begin{split}
    \bar{f_*} &= \mathbf{k}_*^T[K+\sigma_n^2I]^{-1}\mathbf{y} \\
    \mathbb{V}[f_*] &= k(\mathbf{x}_*,\mathbf{x}_*) - \mathbf{k}_*^T[K+\sigma_n^2I]^{-1}\mathbf{k}_*
\end{split}
\end{equation}
$\mathbf{k}_* = k(\mathbf{x}_*,\mathbf{x})$, $K = k(\mathbf{x},\mathbf{x})$. Training a GP is to optimize the hyperparameters $\Theta$ to maximize a marginal likelihood:
\begin{equation}
\text{log}p(f|X) = -\frac{1}{2}\mathbf{y}^T(K+\sigma_n^2I)^{-1}\mathbf{y} - \frac{1}{2}\text{log}|K+\sigma_n^2I|-\frac{n}{2}\text{log}2\pi
\end{equation}

Given the definition of GP in Eq. \ref{eq::gp_def}, the function $g(\cdot)$ can be modeled as a 2D Gaussian Process: 
\begin{equation}
g(R,\beta) \sim \mathcal{GP}(m(\mathbf{x}), k(\mathbf{x},\mathbf{x}')), \mathbf{x} = [R,\beta]
\end{equation}
In this paper, we use the collected experimental data to learn $g(R,\beta)$ through the representation of Gaussian Process.

\begin{table}[b]
  \centering
\caption{Predictive $V^{dis}$ on some typical distances and reflectivities}
  \begin{tabular}{r| p{0.35cm} p{0.35cm} p{0.35cm} p{0.35cm} p{0.35cm} p{0.35cm} |p{0.35cm} p{0.35cm} p{0.35cm}}
    \toprule
    & \multicolumn{6}{c}{$\mathcal{GP}_{diffuse}$ (in meter)} & \multicolumn{3}{|c}{$\mathcal{GP}_{retro}$ (in meter)} \\\hline
   \diagbox[width=4em]{R}{$\beta$} & 1 & 5 & 10 & 30 & 50 & 80  & 120 & 180 & 250 \\\hline
 10m & 80.2 & 51.3 & 38.2  &27.5  & 22.8 & 21.0 & 12.7 & 11.6 & 10.3 \\ 
 15m & 99.3 & 64.2 & 50.1 & 38.9 & 34.8 & 27.6 & 18.5 & 12.2 & 10.9\\ 
 20m & 102.7 & 84.4 & 71.2 & 61.3 & 55.9 & 43.9 & 27.9 & 16.2 & 12.4\\ 
 25m & 134.6 & 104.1 & 94.8 & 79.1 & 66.8 & 52.1 & 34.5 & 21.7 & 13.8\\ 
   \bottomrule
\end{tabular}
\label{tab::re}
\end{table}

\begin{table}[b]
  \centering
\caption{Failure rates of the predictions}
  \begin{tabular}{r| c c}
    \toprule
    &  \multicolumn{2}{c}{Failure rate} \\\hline
   \diagbox[width=4em]{R}{$\beta$} & [0-100] & [100-255]  \\\hline
 0-10m & 2.1\% & 0.7\%  \\ 
 10-15m & 10.3\% & 1.1\% \\ 
 15-20m & 15.1\% & 1.1\% \\ 
 20-25m & 19.5\% & 1.5\%\\\hline
 \textbf{Overall} & 6.5\% &0.8\% \\ 
   \bottomrule
\end{tabular}
\label{tab::failure}
\end{table}

\begin{table*}[t]
  \centering
\caption{Prediction errors of the trained GP model}
  \begin{tabular}{l|cccccccc|c}
    \toprule
   &  \multicolumn{8}{c}{Average predicted $V^{dis}$ errors} \\\hline
  \diagbox[width=9em]{Distance}{Reflectivity} & [0-10) & [10-20) & [20-30) & [30-40) & [40-50) & [50-100) & [100-200) & [200-255] & \textbf{Overall [0-255]}\\\hline
 10m - 15m & 6.23m & 6.21m & 5.89m & 3.32m & 1.78m & 2.18m & 2.11m & 0.55m & \textbf{3.71m} \\
 15m - 20m & 8.21m & 7.17m & 8.10m & 1.48m & 2.36m & 2.22m & 1.41m & 0.41m & \textbf{6.25m}\\ 
 20m - 25m & 22.02m & 16.87m & 12.32m & 10.22m & 5.92m & 12.01m & 4.09m & 1.54m & \textbf{10.72m}\\ 
 25m - 30m & 31.77m & 17.72m & 10.11m & 13.34m & 4.31m & 15.31m & 2.81m & 2.37m & \textbf{12.69m}\\ 
   \bottomrule
\end{tabular}
\label{tab::err}
\end{table*}

\begin{figure*}[t]
\centering
\subfigure[An example of $\mathcal{GP}_{diffuse}$ when $R=10$m, and compared with collected dataset. The blue circles represents the samples utilized to train the GP model. The yellow polyline and shadow region are the predictive mean and 95\% (2$\sigma$) confidence region. Crossings are the collected dataset for verification.]{
\includegraphics[width = 0.45\textwidth]{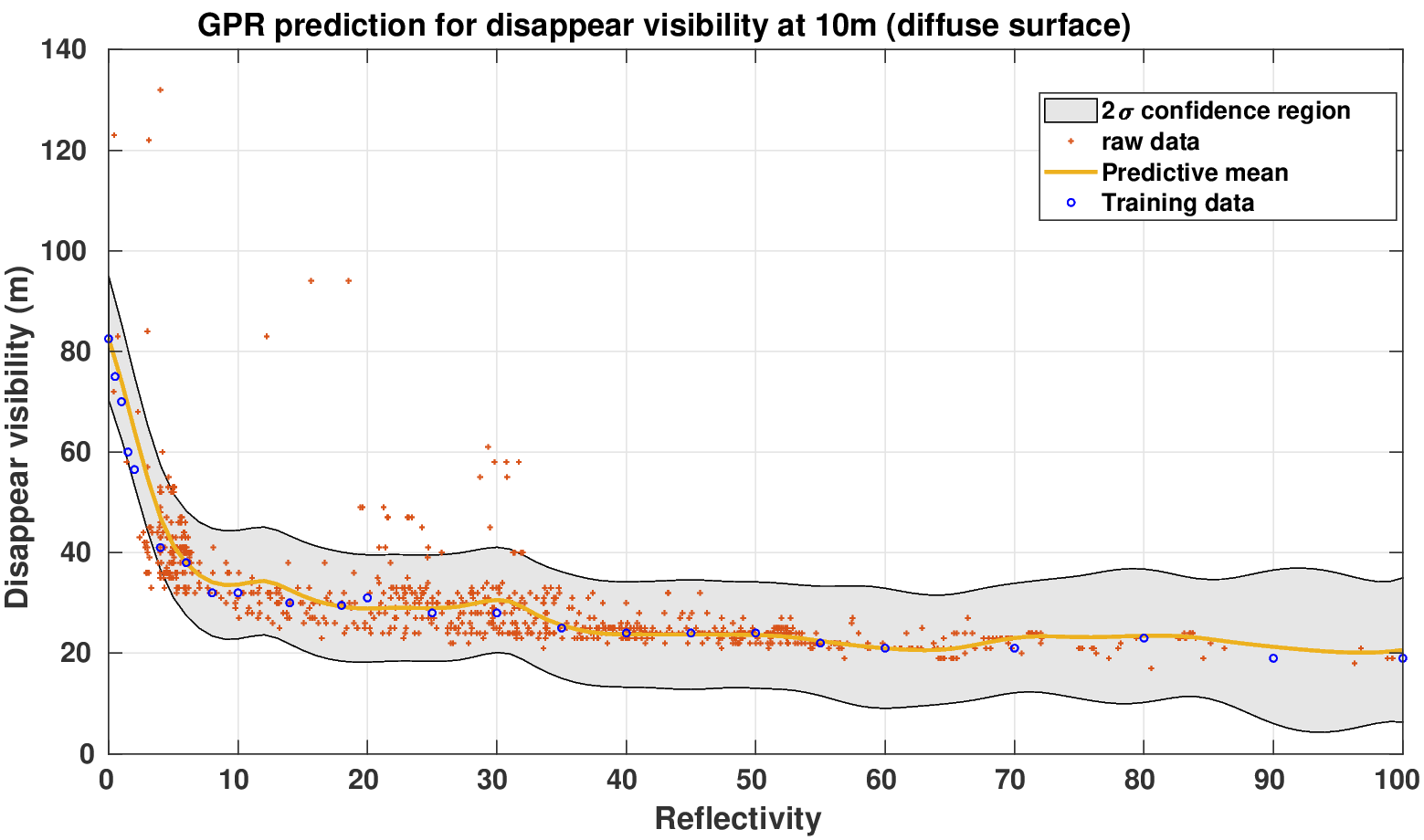}
}
\subfigure[An example of $\mathcal{GP}_{retro}$ when $R=10$m, and compared with collected dataset. The blue circles represents the samples utilized to train the GP model. The yellow polyline and shadow region are the predictive mean and 95\% (2$\sigma$) confidence region. Crossings are the collected dataset for verification. ]{
\includegraphics[width = 0.45\textwidth]{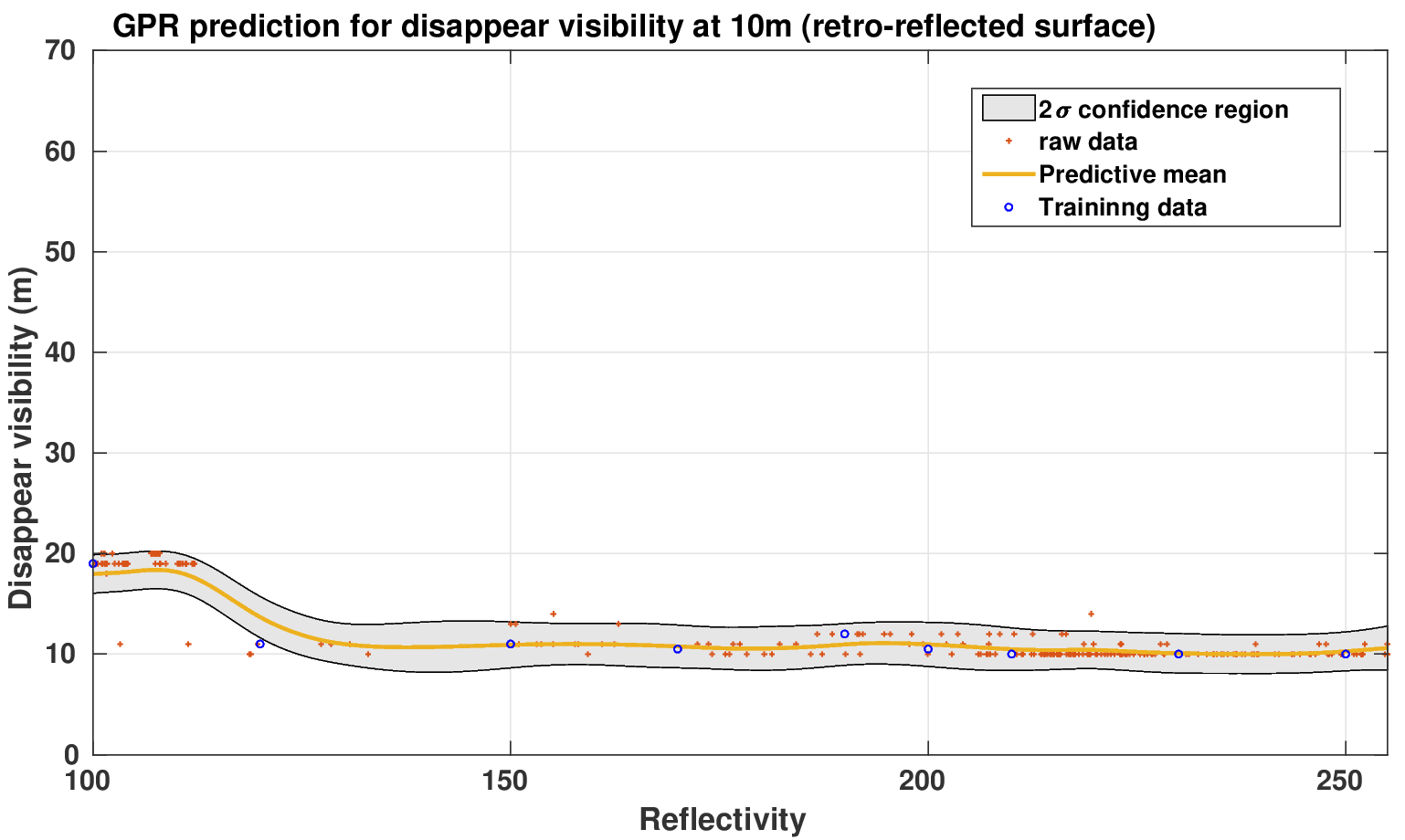}
}
\subfigure[[Predictions of $V^{dis}$ by the $\mathcal{GP}_{diffuse}$ between 10m and 30m.]{
\includegraphics[width = 0.42\textwidth]{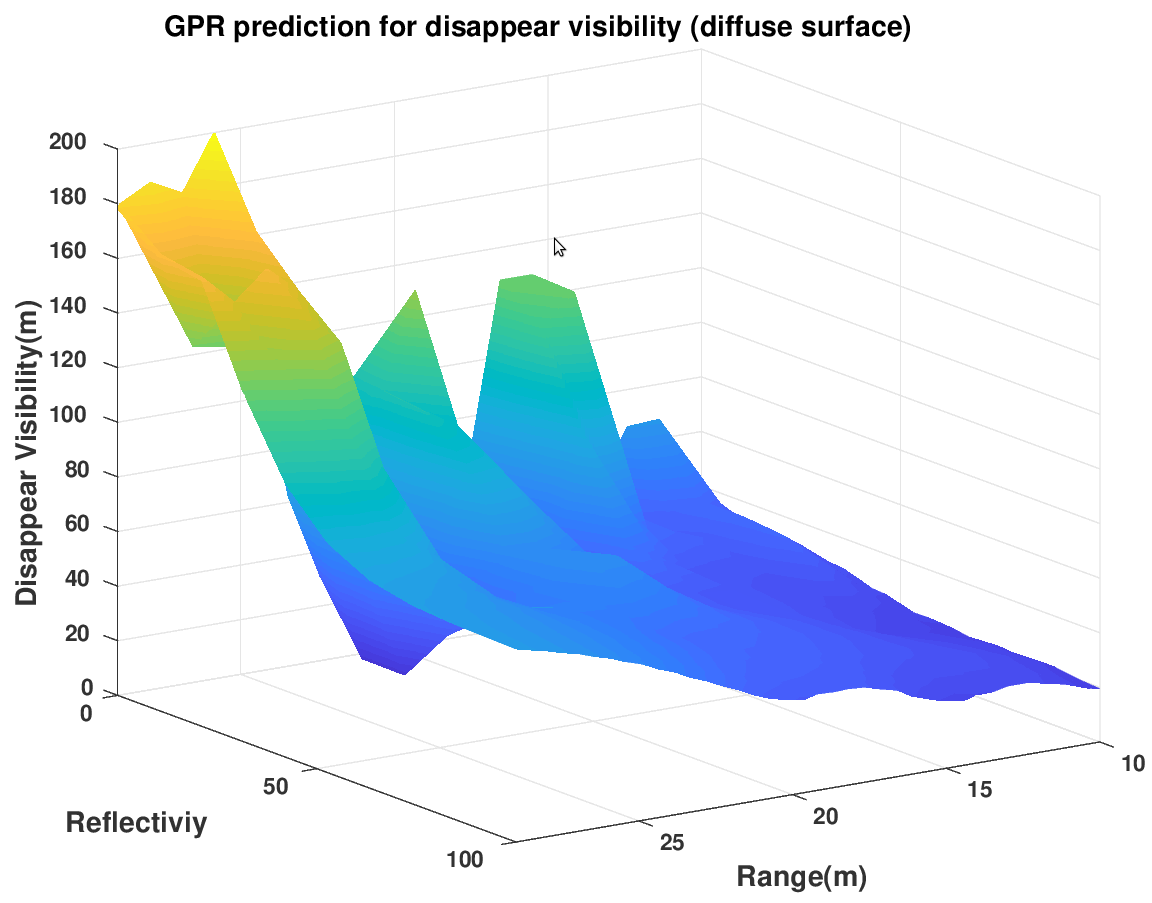}
}
\subfigure[Predictions of $V^{dis}$ by the $\mathcal{GP}_{retro}$ between 10m and 30m.]{
\includegraphics[width = 0.48\textwidth]{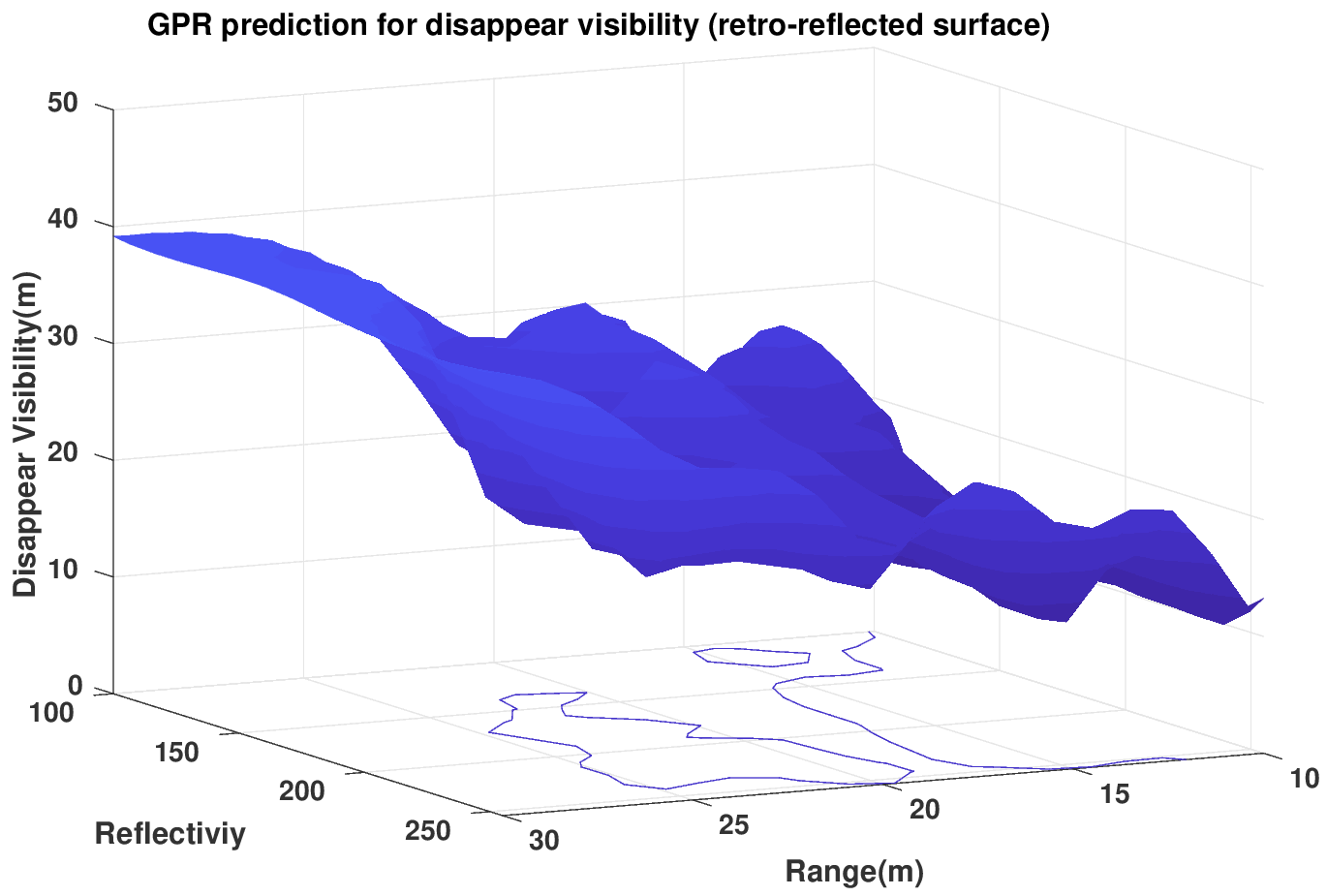}
}
\caption{Trained Gaussian Process and its predictions}
\label{fig::gp_re}
\end{figure*}

\subsubsection{Training GP model}
The average range measures of each second $\bar{r}_{i,j}(\lfloor t \rfloor)$ (in Eq. \ref{eq::average}) of all the lasers within ROI are used for jitter removing. In our test, $3853$ distinctive samples $\{V^{dis}_k | \bar{r}_k, \beta_k\}, k = 1,...,3853$ are collected. The distribution of the collected dataset is not evenly distributed. As shown in Fig. \ref{fig::hist}, there are more samples in short range ($<20m$) and in low reflectivity ($<100$) (Note that we don't have enough samples of relectivities from 60 to 100). There are 565 training data manually selected within $\bar{r} \in[10m,30m]$ and $\beta \in [0,255]$. Because of the difference between diffuse objects ($\beta \in [0,100)$) and retro-reflected objects ($\beta \in [100, 255]$), 2 $\mathcal{GP}$s: $\mathcal{GP}_{diffuse}$ and $\mathcal{GP}_{retro}$ are trained respectively for two different types of objects. The Mat\'ern 3/2 kernel function is chosen for two $\mathcal{GP}$s, because of its finite differentiability that it is able to match physical processes more realistically \cite{ISD1999}:   
\begin{equation}
k(\mathbf{x},\mathbf{x}') = \alpha(1+\frac{\sqrt{3}d}{l})exp(-\frac{\sqrt{3}d}{l}), d = ||\mathbf{x}-\mathbf{x}'||_2
\end{equation}
where $\alpha, l$ are the hyperparameters and $\mathbf{x} = [\bar{r}, \beta]$.  The training is realized by the GPML toolbox. The hyperparameters of the trained two models are: $\mathcal{GP}_{diffuse}$: $\alpha = 0.0678$, $l = 0.1128$, $\sigma_n = 0.0561$, $\mathcal{GP}_{retro}$: $\alpha = 0.0338$, $l = 0.2545$, $\sigma_n = 0.0261$  ($\bar{r}$ and $\beta$ are normalized into [0,1] for training.)

\subsubsection{Results}
The trained GP model is used to predict the $V^{dis}$ for all the reflectivities [0-255] between 10m to 30m, as shown in Fig. \ref{fig::gp_re} (a) - (d). Some typical predictions are shown in Tab. \ref{tab::re}. To evaluate the accuracy of the GP based prediction, we compare the predictions with the real values from the 3853 samples. In the comparison, when the real measured $V^{dis}$ prediction is out of the 95\% confidence region of the prediction, it is classified as a failed prediction. Tab. \ref{tab::failure} demonstrates the failure rates. It shows that our GP model is quite stable for the retro-reflective objects ($\beta > 100$), while for the diffuse reflection targets ($\beta < 100$), the failure rate increases with the distance. While since the dataset contains more samples for short distances, the overall failure rate reaches 6\% for the total dataset.    

For the non-failed predictions, we take the absolute difference between the prediction and real values as the predictive errors. The results are shown in Tab. \ref{tab::err}. Similar to the failure rates, the prediction errors increase along with the distance, particularly for the low reflectivity targets ($\beta<10$).  While the prediction errors for the retro-reflected targets ($\beta>100$) are stable and just around 2 meters. The fact that low reflectivity targets prone to be more disturbed than retro-reflected targets could explain this effect. The reason for the worse performance for long range and low reflective targets is that the ranging process becomes much more noisy than the short and strong reflective targets.

\section{Conclusion}
In this paper, the experimental results of a typical ToF LiDAR under fog environment are demonstrated. Starting from the ranging principle, the factors impacting ToF LiDAR under fog are investigated. Furthermore, we quantitatively evaluate the experimental results and propose a concept of "\textit{disappear visibility}". Following the data-driven principle, we use Gaussian Process to model the distribution of disappear visibility. This method is quite meaningful for evaluating the safety of a LiDAR based autonomous vehicle in fog conditions. 
In the future, we want to test more targets (especially for the ones have reflectivities between 50 to 100) for longer distance as far as more than 100m. Also, more impact factors for the disappear visibility, such as the incident angle between the laser and target's surface, will be considered.  

\section*{Acknowledgment}

This research was funded by the European Union under the H2020 ECSEL Programme as part of the DENSE project (Grant Agreement ID: 692449). DENSE is a joint European project which is sponsored by the European Commission under a joint undertaking. The project was also supported by Groupe RENAULT. We gratefully acknowledge the support from CEREMA and Velodyne.

\bibliographystyle{IEEEtran}
\bibliography{IEEEabrv,uranus}
\end{document}